\documentclass[journal]{IEEEtran}

\ifCLASSINFOpdf
\else
\fi
\usepackage{amssymb}  
\usepackage{graphics} 
\usepackage{epsfig} 
\usepackage{mathptmx} 
\usepackage{times} 
\usepackage{amsmath} 
\usepackage{multicol}
\usepackage{floatrow}
\usepackage{multirow}

\usepackage{xcolor}
\DeclareMathAlphabet{\mathcal}{OMS}{cmsy}{b}{n}
\DeclareMathAlphabet{\mathcal}{OMS}{cmsy}{m}{n}
\newtheorem{theorem}{\indent Theorem}
\newtheorem{lemma}{\indent Lemma}

\newtheorem{definition}{\indent Definition}
\newtheorem{example}{\indent Example}
\newtheorem{remark}{\indent Remark}
\newtheorem{criterion}{\indent Criterion}
\newtheorem{problem}{\indent Problem}

\newtheorem{property}{\indent Property}
\begin{document}


\title{Quantum Hamiltonian Identifiability via a Similarity Transformation Approach and Beyond} 
\author{Yuanlong~Wang,
~Daoyi~Dong,
Akira~Sone,
Ian~R.~Petersen,
Hidehiro~Yonezawa,
Paola~Cappellaro

\thanks{This work was supported in part by the Australian Research Council's Discovery Projects Funding Scheme under Project DP130101658, in part by the Laureate Fellowship FL110100020, in part by the Air Force Office of Scientific Research under Agreement FA2386-16-1-4065, in part by the Centres of Excellence CE170100012, in part by the U.S. Army Research Office through Grants No. W911NF-11-1-0400 and No. W911NF-15-1-0548, and in part by the NSF Grant No. PHY0551153. (\textit{Corresponding authors: Daoyi Dong and Paola Cappellaro.})}
\thanks{Y. Wang and H. Yonezawa are with the School of Engineering and Information Technology, University of New South Wales, Canberra, ACT 2600, Australia, and also with the Centre for Quantum Computation and Communication Technology, Australian Research Council, Canberra, ACT 2600, Australia (e-mail: yuanlong.wang.qc@gmail.com; h.yonezawa@adfa.edu.au).}%
\thanks{D. Dong is with the School of Engineering and Information Technology, University of New South Wales, Canberra, ACT 2600, Australia (e-mail: daoyidong@gmail.com).}
\thanks{A. Sone and P. Cappellaro are with the Research Laboratory of Electronics and Department of Nuclear Science and Engineering, Massachusetts Institute of Technology, Cambridge, Massachusetts 02139, USA (e-mail: asone@mit.edu; pcappell@mit.edu).}
\thanks{I. R. Petersen is with the Research School of Engineering, Australian National University, Canberra, ACT 2601, Australia (e-mail: i.r.petersen@gmail.com).}
}
\markboth{Quantum system identifiability}
{Shell \MakeLowercase{\textit{et al.}}: Bare Demo of IEEEtran.cls for IEEE Journals}
\maketitle
\begin{abstract}                          
The identifiability of a system is concerned with whether the unknown parameters in the system can be uniquely determined with all the possible data generated by a certain experimental setting. A test of quantum Hamiltonian identifiability is an important tool to save time and cost when exploring the identification capability of quantum probes and experimentally implementing quantum identification schemes. In this paper, we generalize the identifiability test based on the Similarity Transformation Approach (STA) in classical control theory and extend it to the domain of quantum Hamiltonian identification. We employ STA to prove the identifiability of spin-1/2 chain systems with arbitrary dimension assisted by single-qubit probes. We further extend the traditional STA method by proposing a Structure Preserving Transformation (SPT) method for non-minimal systems. We use the SPT method to introduce an indicator for the existence of economic quantum Hamiltonian identification algorithms, whose computational complexity directly depends on the number of unknown parameters (which could be much smaller than the system dimension). Finally, we give an example of such an economic Hamiltonian identification algorithm and perform simulations to demonstrate its effectiveness.
\end{abstract}

\begin{IEEEkeywords}                           
Quantum system; Hamiltonian identifiability; quantum Hamiltonian identification; similarity transformation approach.
\end{IEEEkeywords}                             

\IEEEpeerreviewmaketitle

\section{INTRODUCTION}\label{Secintro}
\IEEEPARstart{T}{here} is growing interest in quantum system research, aiming to develop advanced technology including quantum computation, quantum communication \cite{Nielsen and Chuang 2000} and quantum sensing \cite{quantum sensing}. Before exploiting a quantum system as a quantum device, it is usually necessary to estimate the state and identify key variables of the system \cite{paris 2004}-\cite{sone 2018}. The Hamiltonian is a fundamental quantity that governs the evolution of a quantum system. Hamiltonian identification is thus critical for tasks such as calibrating quantum devices \cite{wang 2017} and characterizing quantum channels \cite{devitt 2006,zhang 2015}.

Before performing identification experiments, a natural question arises: is the available data from a given experimental setting enough to identify (or determine) all the desired parameters in the Hamiltonian? In this paper, we refer to such a problem as Hamiltonian identifiability. The solution to this problem is fundamental and necessary for designing experiments, and also gives us insights into the information extraction capability of certain probe systems.

There are several existing approaches to investigating the problem of quantum system identification \cite{fu 2016}-\cite{shu 2016}. For example, Ref. \cite{burgarth 2012} proved that controllable quantum systems are indistinguishable if and only if they are related through a unitary transformation, which can be developed as an identifiability method for controllable systems. The identifiability problem for a Hamiltonian corresponding to a dipole moment was investigated in \cite{bris 2007}. The identification problem of spin chains has been extensively investigated in e.g., \cite{franco 2011}-\cite{burgarth 2017b}. Ref. \cite{guta 2016} presented identifiable conditions for parameters in passive linear quantum systems, and further disposed of the requirement of ``passive" in \cite{levitt 2017}. Control signals to enhance the observability of the quantum dipole moment matrix were introduced in \cite{leghtas 2012}. Zhang and Sarovar \cite{zhang 2014} proposed a Hamiltonian identification method based on measurement time traces. Sone and Cappellaro \cite{sone 2017} employed Gr\"{o}bner basis to test the Hamiltonian identifiability of spin-1/2 systems, and their method is also applicable to general finite-dimension systems.

We assume the dimension \cite{sone dimension} and structure (e.g., the coupling types) \cite{kato 2014} of the Hamiltonian is already determined, and the task is to identify unknown parameters in the Hamiltonian. It is natural to resort to identifiability test methods in classical (non-quantum) control field to tackle the quantum Hamiltonian identifiability problem. Common classical methods include the Laplace transform approach \cite{bellman 1970}, the Taylor series expansion approach \cite{pohjanpalo 1978} and the Similarity Transformation Approach (STA) \cite{travis 1981}-\cite{vajda 1989}. For a review, see \cite{godfrey 1987}-\cite{book 2011}. The main idea of the Laplace transform approach is to determine the number of solutions of the multivariate equations composed by coefficients of the transfer function. In contrast, the STA method transforms the identifiability problem into finding the existence of unequal solutions of similarity equations generated by a minimal system's equivalent realizations, thus providing a chance to avoid directly solving multivariate polynomial equations, a considerable advantage in the case of high-dimension or incomplete prior information. In this paper, we extend the STA method to quantum Hamiltonian identifiability. We generalize and improve STA-based identifiability criteria, which are applicable to both classical control and quantum identification domains. We employ the STA method to analyze all physical cases in \cite{sone 2017} and present proofs for the associated identifiability conclusions.

We further propose a Structure Preserving Transformation (SPT) method for the STA-based identifiability analysis in non-minimal systems. In classical control, when faced with non-minimal systems, one usually prefers to change the system settings such that it becomes minimal. In other words, the original settings are abandoned. This indirect solution is not applicable when the experimental settings are difficult to change or when we only expect to explore the information extraction capability of some particular physical probe systems. However, the SPT method provides a chance to preserve most of the system key properties after transformations while still performing identifiability analysis on its minimal subsystem. Hence, we employ the SPT method to prove that it is always possible to estimate one unknown parameter in the system matrix using a specifically designed experimental setting. This conclusion serves as an indicator for the existence of ``economic" quantum Hamiltonian identification algorithms, whose computational complexity directly depends on the number of unknown parameters.

As an example, we provide a specific economic identification algorithm. The computational complexity $O(\mathcal{M}^2+q\mathcal{M}\mathcal{N})$ only depends on the number of unknown parameters $\mathcal{M}$ and data length $\mathcal{N}$ ($q$ is a variable not larger than $\mathcal{N}$). Therefore, for physical systems with a small number of unknown parameters in the Hamiltonian, this identification algorithm can be efficient.

The main contributions of this paper are summarized as follows,

\begin{itemize}
\item The identifiability test method based on Similarity Transformation Approach (STA) is generalized and extended to the quantum Hamiltonian identifiability problem. Identifiability test criteria are improved and the analysis method for non-minimal systems based on Structure Preserving Transformation (SPT) is proposed.
\item Based on the STA method, three physical cases in \cite{sone 2017} are analyzed and the identifiability conclusions are proved for the systems with arbitrary dimension.
\item To analyze general non-minimal systems, an SPT method is developed to present an indicator for the existence of economic Hamiltonian identification algorithms, which have computational complexity directly depending on the number of unknown parameters. One example of such algorithms is then presented.
\end{itemize}

The structure of this paper is as follows. In Sec. \ref{sec1} we present some preliminaries, formulate the identifiability problem and briefly introduce the classical method based on the Laplace transform approach. Sec. \ref{sec2} presents the identifiability test method employing STA. Based on the STA method, we present the identifiability proof for two spin models, the exchange model without and with transverse field in Sec. \ref{seccase1} and \ref{seccase2}, respectively. In Sec. \ref{secevi} we employ the SPT method to present an indicator for the existence of economic quantum Hamiltonian identification algorithms and also give a concrete example of developing such an algorithm. Sec. \ref{secfinal} concludes the paper.

Notation: $*$ denotes an indeterminate variable or matrix. For a matrix $A$, $A_{\sigma i}$ and $A_{j\sigma}$ denote its $i$-th column and $j$-th row, respectively. $\mathbb{R}$ and $\mathbb{C}$ are real and complex domains, respectively. $\otimes$ is the tensor product. $\text{vec}(\cdot)$ is the column vectorization function. $\lambda_i{(A)}$ is the $i$th eigenvalue of $A$ and $\Lambda(A)$ is the set of all the eigenvalues of $A$ (repeated eigenvalues appear multiple times in $\Lambda(A)$). $||\cdot||$ is the Frobenius norm. $\delta$ is the Dirac delta function or the Kronecker delta function, in the continuous or discrete sense, respectively. $\hat{x}$ denotes the estimation value of the true value $x$. $\lfloor x\rfloor$ returns the largest integer that is not larger than $x$.

\section{PRELIMINARIES AND PROBLEM FORMULATION}\label{sec1}

\subsection{Quantum state and measurement}

The state of a quantum system is represented by a complex Hermitian matrix $\rho$ in a Hilbert space and its dynamics are described by the Liouville-von Neumann equation
\begin{equation}\label{eq0}
\dot{\rho}=-\text{i}[H,\rho],
\end{equation}
where $\text{i}=\sqrt{-1}$, $H$ is the system Hamiltonian, $[A,B]=AB-BA$ is the commutator and we set $\hbar=1$ using atomic units in this paper. $\rho$ is a positive semidefinite matrix satisfying $\text{Tr}(\rho)=1$.

To extract information from a quantum state, it is normally necessary to perform a positive-operator valued measurement (POVM), which is a set $\{M_i\}$, where all the elements are Hermitian positive semidefinite matrices and $\sum_iM_i=I$. When a set $\{M_i\}$ of POVM is performed, the probability of outcome $i$ occurring is determined by the Born Rule, $p_i=\text{Tr}(\rho M_i)$. The data in actual experiments are the approximation values of $p_i$.

\subsection{Problem formulation of Hamiltonian identifiability and identification}\label{subsec2}
We first rephrase the framework in \cite{zhang 2014} to recast the problem of Hamiltonian identification as a linear system identification problem. Let $H$ be the $d$-dimensional Hamiltonian to be identified, which can be parametrized as
\begin{equation}\label{eq1}
H=\sum_{m=1}^{\mathcal{M}} a_m(\mathbf\theta)H_m,
\end{equation}
where $\mathbf{\theta}=(\theta_1,...,\theta_{\mathcal{M}})^T$ is a vector consisting of all the unknown parameters, ${\mathcal{M}}$ is the number of unknown parameters, $a_m$ are known functions of $\mathbf{\theta}$ and $H_m$ are known Hermitian matrices (also called basis matrices). Let $\mathfrak{su}(d)$ denote the Lie algebra consisting of all $d\times d$ skew-Hermitian traceless matrices. Then $\{H_m\}$ can be chosen as an orthonormal basis of $\mathfrak{su}(d)$, where the inner product is defined as $\langle \text{i}H_m,\text{i}H_n\rangle=\text{Tr}(H_m^\dagger H_n)$. The traceless assumption is reasonable because $H$ has an intrinsic degree of freedom (see \cite{my 2016} for details).

Let $S_{jkl}$ be the structure constants of $\mathfrak{su}(d)$, which satisfy
\begin{equation}
[\text{i}H_j,\text{i}H_k]=\sum_{l=1}^{d^2-1}S_{jkl}(\text{i}H_l),
\end{equation}
where $j,k=1,...,d^2-1$. If $H_k$ is the observable, then the experimental data is obtained from Born's rule
\begin{equation}\label{eq4}
x_k=\text{Tr}(H_k\rho).
\end{equation}
The identifiability is determined by the system structure. Hence, it is usually assumed that there are no imperfections in the available experimental data, which is the reason we identify theoretical values with practical data in (\ref{eq4}).

From (\ref{eq0})-(\ref{eq4}) we have
\begin{equation}\label{eq5}
\dot{x_k}=\sum_{l=1}^{d^2-1}(\sum_{m=1}^{{\mathcal{M}}}S_{mkl}a_m(\mathbf\theta))x_l.
\end{equation}

If we directly rewrite (\ref{eq5}) into a matrix form, the dimension of the system matrix would be $d^2-1$, which is large for multi-qubit systems. To reduce the dimension, first consider the operators $O_i$ that we can directly measure in practice. We expand $O_i$ as $O_i=\sum_j o_j H_j$, and collect all the $H_j$ that appear in the expansion of $O_i$ as $\mathbb{M}=\{H_{v_1},...,H_{v_p}\}$. Also, we collect all the $H_j$ that appear in the expansion of $H$ as $\mathbb{L}=\{H_m\}_{m=1}^{{\mathcal{M}}}$. Define an iterative procedure as $$G_0=\mathbb{M}, \ \ G_i=\{G_{i-1},\mathbb{L}\}\cup G_{i-1},$$ where $\{G_{i-1},\mathbb{L}\}\triangleq \{H_j|\text{Tr}(H_j^\dagger[g,h])\neq 0, g\in G_{i-1}, h\in \mathbb{L}\}$. This iteration will terminate at a maximal set $\bar G$ (called the \textit{accessible set}) because $\mathfrak{su}(d)$ is finite. We collect all the $x_i$ with $H_i\in\bar G$ in a vector $\mathbf{x}$ of dimension $n$, and its dynamics satisfy the linear system equation
\begin{equation}\label{eq7}
\mathbf{\dot{x}}=A\mathbf{x}.
\end{equation}
The elements in $A$ are the coefficients in (\ref{eq5}), which are linear combinations of $a_m(\mathbf{\theta})$. For some types of physical systems, the dimension $n$ can be much smaller than $d^2-1$. $A$ is real and antisymmetric due to the antisymmetry of the structure constants. The output data can be denoted as
\begin{equation}\label{eq8}
\mathbf{y}=C\mathbf{x},
\end{equation}
where $C$ is a known matrix. Therefore, the quantum Hamiltonian identification problem can be established as follows:

\begin{problem}
Given the system matrix $A=A(\mathbf\theta)$, initial state $\mathbf{x}(0)=\mathbf{x}_0$ and observation matrix $C$, design an algorithm to obtain an estimate $\mathbf{\hat\theta}$ of $\mathbf{\theta}$ from measurement data $\mathbf{\hat y}$.
\end{problem}

Before designing specific identification algorithms, a natural question arises: for a system $A$, can we uniquely determine the unknown parameters, based on a given experimental setup (i.e., $\mathbf{x}_0$ and $C$)? If not, then it may be required to redesign the experimental setup before starting the experiment. This is especially significant for quantum system identification, since implementing quantum experiments is usually expensive. This induces the problem of identifiability. Denote $\mathbf{\theta}$ the true value of the unknown parameter vector to be identified. Assume that the system under consideration has some parametric model structure with output data $\mathcal{S}(\mathbf\theta)$, for a given experimental setup. The equation
\begin{equation}\label{eqid}
\mathcal{S}(\mathbf{\theta})=\mathcal{S}(\mathbf{\theta'})
\end{equation}
means that the model with parameter set $\mathbf{\theta'}$ outputs \textit{exactly} the same data as the model with parameter set $\mathbf{\theta}$. Identifiability then depends on the number of solutions to (\ref{eqid}) for $\mathbf{\theta'}$. We use the following definition from \cite{book 2011}:

\begin{definition}\cite{book 2011}\label{def1}
The model $\mathcal{S}$ is \textit{structurally globally identifiable} (abbreviated as \textit{identifiable} in the rest of this paper), if for almost any value of $\mathbf{\theta}$, (\ref{eqid}) has only one solution $\mathbf{\theta'}=\mathbf{\theta}$.
\end{definition}

Definition \ref{def1} is in essence the same as the definition of identifiability in \cite{sone 2017}. It is necessary to ensure identifiability holds for almost any value of the parameters because the number of solutions to (\ref{eqid}) might change for some particular values of $\mathbf{\theta}$, which are called \textit{atypical} cases (to be illustrated later). Also, identifiability is determined by the system structure. Hence, we do not consider noise or uncertainty in the experimental data. A trivial necessary condition for a parameter to be identifiable is that it should appear in the system model $\mathcal{S}$, and in the following we only focus on this class of parameters.

\subsection{The Laplace transform approach and atypical cases}\label{laplace}
One of the most intuitive ways to solve identifiability problems is through the Laplace transform, which is also helpful in understanding concepts like \textit{atypical} cases. Hence, we first briefly introduce the Laplace transform approach \cite{book 2011}. Consider the following standard MIMO linear system with zero initial condition:
\begin{equation}\label{eq10}
\left\{
\begin{array}{rl}
\mathbf{\dot{x}}&=A(\mathbf\theta)\mathbf{x}+B(\mathbf\theta)\mathbf{u},\ \ \mathbf{x}(0)=\mathbf{0},\\
\mathbf{y}&=C(\mathbf\theta)\mathbf{x}+D(\mathbf\theta)\mathbf{u}.\\
\end{array}
\right.
\end{equation}
Throughout this paper we use 4-tuples $\Sigma=(A,B,C,D)$ to denote linear systems with the form of (\ref{eq10}). The Laplace transform solution to (\ref{eq10}) is $$\mathbf{Y}(s,\mathbf\theta)=\mathbf{T}(s,\mathbf\theta)\mathbf{U}(s),$$ where the transfer function matrix is $\mathbf{T}(s,\theta)=C(\mathbf\theta)[sI-A(\mathbf\theta)]^{-1}B(\mathbf\theta)+D(\theta)$. In the frequency domain, (\ref{eqid}) is now $$\mathbf{T}(s,\mathbf{\theta})\mathbf{U}(s)=\mathbf{T}(s,\mathbf{\theta'})\mathbf{U}(s).$$ By cancelling $\mathbf{U}(s)$, (\ref{eqid}) is equivalent to
\begin{equation}\label{eq12}
\mathbf{T}(s,\mathbf{\theta})=\mathbf{T}(s,\mathbf{\theta'}),\ \ \forall s.
\end{equation}
Hence, the transfer function is exactly a tool to characterize identifiability. By writing (\ref{eq12}) in a canonical form (e.g., transforming the numerators and denominators into monic polynomials) and equating coefficients on both sides of (\ref{eq12}), one obtain a series of algebraic equations in $\mathbf{\theta}$ and $\mathbf{\theta'}$. If for almost any value of $\mathbf{\theta}$, the solutions always satisfy $\mathbf{\theta'}=\mathbf{\theta}$, then the system is identifiable. In order to investigate identifiability, Sone and Cappellaro \cite{sone 2017} employed Gr\"{o}bner basis to determine the conditions of identifiability. By directly solving (\ref{eq12}) where the RHS is replaced by a specific transfer function reconstructed from experimental data, one can develop algorithms like that in \cite{zhang 2014} to identify the Hamiltonian.

The following property of the transfer function will be frequently used in the sequel:
\begin{property}\label{property1}
When a system undergoes a similarity transformation $\mathbf{x'}=P\mathbf{x}$ where $P$ is a nonsingular matrix, the transfer function remains the same, and thus the identifiability does not change.
\end{property}

We specifically illustrate \textit{atypical} cases and \textit{hypersurfaces}. Assume that the number of unknown parameters is ${\mathcal{M}}$ and we have no prior knowledge of the true values, which indicates the candidate space for the parameters is $\mathbb{R}^{\mathcal{M}}$. A hypersurface is a manifold or an algebraic variety with dimension ${\mathcal{M}}-1$, and it is usually obtained by adding an extra polynomial equation about the unknown parameters. Hypersurface sets have Lebesgue measure zero and they can thus be neglected in practice. Atypical cases are subsets of hypersurfaces. Hence, analysis on atypical cases can also be omitted. When the complement of a hypersurface is open and dense in $\mathbb{R}^{\mathcal{M}}$ and has full measure, it is often called a \textit{generic} set \cite{Sontag 2002}. For strictness, the phrase ``almost always" is usually employed to indicate that atypical cases have already been neglected. We give an example of atypical cases from the point of view of transfer functions like Example 3.1 in \cite{book 2011}. Consider a system with unknown parameters $\theta_1$ and $\theta_2$ and the transfer function
\begin{equation}\label{equationadd03}
\mathbf{T}(s,\mathbf{\theta})= \frac{\theta_1}{s+\theta_1+\theta_2}.
\end{equation}
The algebraic equations from (\ref{eq12}) are thus $\theta_1=\theta_1'$ and $\theta_1+\theta_2= \theta_1'+\theta_2'$. Therefore, the system (\ref{equationadd03}) is generally identifiable, except the case of $\theta_1=0$ which leads to a zero transfer function and erases all the information about $\theta_2$. Since $\theta_1=0$ is an atypical case, we can omit it and conclude that this system is (almost always) identifiable. In the rest of this paper we omit ``almost always" if there is no ambiguity.


\section{HAMILTONIAN IDENTIFIABILITY VIA THE SIMILARITY TRANSFORMATION APPROACH}\label{sec2}
\subsection{General procedures for minimal systems}\label{sec2a}
Strictly speaking, the word ``minimal" is used to describe system realizations that are both controllable and observable. In this paper, we call a system ``minimal" if it is both controllable and observable.

Let $\mathbf{\theta}$ be the true value generating the system (\ref{eq10}). Suppose that there is an alternative value $\mathbf{\theta'}$ generating the same output data. Then $\mathbf{\theta'}$ gives an alternative realization:
\begin{equation}\label{eq13}
\left\{
\begin{array}{rl}
\mathbf{\dot{x}'}&=A(\mathbf\theta')\mathbf{x}'+B(\mathbf\theta')\mathbf{u},\ \ \mathbf{x}'(0)=\mathbf{0},\\
\mathbf{y}&=C(\mathbf\theta')\mathbf{x}'+D(\mathbf\theta')\mathbf{u}.\\
\end{array}
\right.
\end{equation}
Suppose that the system realization (\ref{eq10}) is minimal, then (\ref{eq13}) is also minimal since they have the same dimension. From Kalman's algebraic equivalence theorem \cite{kalman 1963}, minimal realizations of a transfer function are equivalent; i.e., they are related by a similarity transformation:
\begin{equation}\label{eq14}
\left\{
\begin{array}{rl}
&A(\mathbf\theta)=S^{-1}A(\mathbf{\theta'})S,\\
&B(\mathbf{\theta})=S^{-1}B(\mathbf{\theta'}), \\
&C(\mathbf{\theta})=C(\mathbf{\theta'})S,\\
&D(\mathbf{\theta})=D(\mathbf{\theta'}),\\
\end{array}\right.
\end{equation}
where $S$ is an invertible matrix. We call equations (\ref{eq14}) the \textit{STA equations}. We take $S$, $\mathbf\theta$ and $\mathbf{\theta'}$ as unknown variables and search for their solution. The solvability of (\ref{eq14}) can be guaranteed because it always has a trivial solution $S=I$ and $\mathbf{\theta}=\mathbf{\theta'}$. If all the solutions satisfy $\mathbf{\theta}=\mathbf{\theta'}$, then the system (\ref{eq10}) is identifiable. Otherwise it is unidentifiable. In cases when the signs of $\mathbf{\theta}$ are not considered, one can check whether all the solutions to the STA equations satisfy $|\mathbf\theta|=|\mathbf{\theta'}|$ to determine the identifiability.

\subsection{Non-minimal systems}\label{sec2b}

\begin{figure}
\centering
\includegraphics[width=3in]{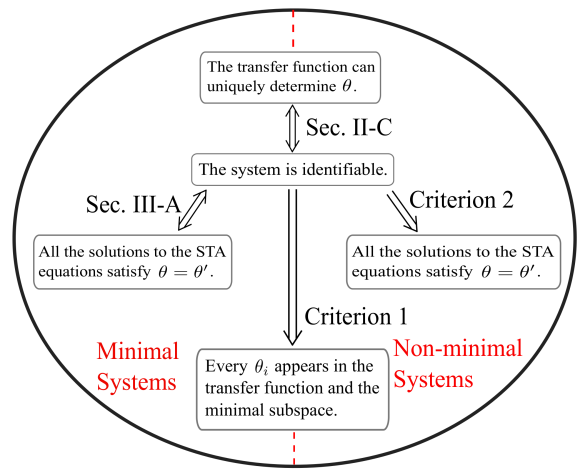}
\centering{\caption{Relationships between identifiability criteria.}\label{fig1}}
\end{figure}

If the system is not minimal, Kalman's algebraic equivalence theorem (and hence the STA equations) can only be applied to the controllable and observable part of the system. If one ignores whether the system is minimal or not and directly employs the solution to the STA equations to test the identifiability, an incorrect conclusion might be obtained. For example, consider the following 2-dimensional system:
\begin{example}\label{example1}
\begin{equation}\label{eq26}
\left\{
\begin{array}{rl}
\mathbf{\dot{x}}&=\left(\begin{matrix}\theta_1& 0 \\ 0 &\theta_2\end{matrix}\right)\mathbf{x}+\left(\begin{matrix}1 \\ 0\end{matrix}\right){u},\ \ \mathbf{x}(0)=\mathbf{0},\\
{y}&=(1\ \ 0)\mathbf{x}.\\
\end{array}
\right.
\end{equation}
\end{example}
This system (\ref{eq26}) is uncontrollable and unobservable. If one directly solves the STA equations, the conclusion is that it is identifiable. However, since the output $y$ never contains any information about $x_2$, which evolves independently as $\dot{x}_2=\theta_2x_2$, $\theta_2$ is in fact unidentifiable.

The fact that (\ref{eq12}) is equivalent to (\ref{eqid}) in Sec. \ref{laplace} means a linear system's identifiability is uniquely and completely determined by its transfer function. Therefore, unlike in the situation using STA, non-minimal systems do not introduce extra requirements in the Laplace transform approach.

Regardless of controllability or observability, the transfer function of a system remains the same under similarity transformation. Therefore, for uncontrollable or unobservable systems, the solution using STA is \cite{walter 1981}: (i) perform Kalman decomposition and obtain the controllable and observable (minimal) subsystem; (ii) write down the STA equations for the minimal subsystem; (iii) the original system is identifiable if and only if the solutions to the STA equations in (ii) all satisfy $\mathbf{\theta}=\mathbf{\theta'}$.

For Example \ref{example1}, (\ref{eq26}) is already in the Kalman canonical form and the minimal subsystem is $\dot{x}_1=\theta_1x_1+u$, $y=x_1$. Hence, $\theta_1$ is identifiable and $\theta_2$ is unidentifiable. This example also implies the following identifiability Criterion \ref{cri1}, which corresponds to the fact in \cite{sone 2017} that the parameters that do not appear in the transfer function are unidentifiable.
\begin{criterion}\label{cri1}
Suppose a system is non-minimal. Perform Kalman decomposition to obtain its minimal subsystem and non-minimal subsystem. The unknown parameters that do not appear in the minimal subsystem are unidentifiable.
\end{criterion}

For a non-minimal system, even if all the unknown parameters appear in the minimal subsystem and the STA equations for the original system (rather than the minimal subsystem) exclude the solutions $\mathbf{\theta}\neq\mathbf{\theta'}$, it is not sufficient for guaranteeing the identifiability of the original system. A straightforward example can be obtained by substituting $\theta_1$ and $\theta_2$ in Example \ref{example1} with $\theta_1+\theta_2$ and $\theta_1-\theta_2$, respectively.

Although it is necessary to analyze the minimality before solving the STA equations in most situations, we find a shortcut for some special cases.

\begin{criterion}\label{cri2}
If the STA equations for a system have a (non-atypical) solution $\mathbf{\theta_0}\neq\mathbf{\theta_0}'$, the system is unidentifiable regardless of whether it is minimal or not.
\end{criterion}

For the proof of Criterion \ref{cri2}, we consider two specific realizations $(A(\mathbf{\theta_0}),B(\mathbf{\theta_0}),C(\mathbf{\theta_0}),D(\mathbf{\theta_0}))$ and $(A(\mathbf{\theta_0}'),B(\mathbf{\theta_0}'),C(\mathbf{\theta_0}'),D(\mathbf{\theta_0}'))$ for the system. According to the form of STA equations (\ref{eq14}), these two different (possibly non-minimal) realizations are related by a similarity transformation. Using Property \ref{property1} they result in the same transfer function. Therefore, different system parameters are generating the same system model. This means the system must be unidentifiable, which proves Criterion \ref{cri2}.

As pointed out in \cite{distefano 1977}, the controllability and observability properties are neither sufficient nor necessary for identifiability. Example \ref{example1} has shown that non-minimal systems may be unidentifiable. If one replaces $\theta_2$ in the system matrix of (\ref{eq26}) with $\theta_1$, then the system becomes identifiable, which indicates non-minimal systems can also be identifiable.

In Fig. \ref{fig1}, we summarize all the results of Sec. \ref{sec2a} and \ref{sec2b}. Note that for non-minimal systems Criterion \ref{cri2} is necessary but not sufficient, different from the case for minimal systems.

\subsection{Structure Preserving Transformation method}
Structure Preserving Transformation (SPT) method is an idea we develop for identifiability analysis in non-minimal systems. Suppose there is a non-minimal system $\Sigma=(A,B,C,D)$ with state vector $\mathbf{x}$. If Criterion \ref{cri2} fails, traditionally we have to perform Kalman decomposition. We let $\mathbf{\bar{x}}=P\mathbf{x}$ such that the equivalent system $\bar{\Sigma}=(\bar{A},\bar{B},\bar{C},\bar{D})$ has the Kalman canonical form. Then, we employ the STA equations for its minimal subsystem $\bar{\Sigma}_1=(\bar{A}_1,\bar{B}_1,\bar{C}_1,\bar{D}_1)$, with the corresponding state vector $\mathbf{\bar{x}_1}$ having a dimension smaller than $\mathbf{x}$.

Quantum systems usually generate clear structure properties in $A$. These structure properties may be completely disguised in the system $\bar{\Sigma}$, making the STA equations difficult to solve. This problem is seldom investigated in classical control theory, because classically one prefers to change the system structure $(A,B,C,D)$ so that the system becomes minimal when faced with such problems. On the contrary, quantum research sometimes investigates the physical capability of a certain fixed system setting and the initial quantum system states or the observables may be difficult to change. Therefore, changing $(A,B,C,D)$ may not be practical. How can we keep (some of) the structure properties of the original system $\Sigma$ and meanwhile perform STA analysis?

The idea of SPT is to further perform a similarity transformation on $\bar{\Sigma}$ to recover (some of) the structure properties of $\Sigma$, meanwhile preserving the canonically decomposed form. To do this, we let $\mathbf{\tilde{x}}=\text{diag}(\tilde{P}^{-1},I)\mathbf{\bar{x}}$ and obtain a system $\tilde{\Sigma}=(\tilde{A},\tilde{B},\tilde{C},\tilde{D})$, where $\tilde{P}^{-1}$ acts only on the minimal subsystem $\bar{\Sigma}_1$. Since the second transformation $\text{diag}(\tilde{P}^{-1},I)$ is block-diagonal, $\tilde{\Sigma}$ is still in the Kalman canonical form, and the matrices $(\tilde{A}_1,\tilde{B}_1,\tilde{C}_1,\tilde{D}_1)$ are submatrices of those in $\tilde{\Sigma}$, respectively. If $\tilde{P}$ is close to $P$ (in the form/appearance, not in norm), or $\tilde{P}^{-1}$ is close to $P^{-1}$, then we are likely to regain an $\tilde{A}_1$ similar to $A$, thus recovering key structure properties. Then we solve the STA equations for the minimal subsystem $\tilde{\Sigma}_1$ to determine the identifiability.

In the SPT method, $\tilde{P}$ can never be exactly equal to $P$, because their dimensions are different. The choice of $\tilde{P}$ is not unique and should depend on specific problems. One common choice is to let $\tilde{P}$ be a submatrix of $P$. An example using the SPT method is provided in Sec. \ref{secevi1}.

\subsection{Quantum Hamiltonian identifiability via STA}

We clarify several points when using STA for analyzing Hamiltonian identifiability. For simplicity we only consider single input Hamiltonian systems (i.e., the state variable $\mathbf{x}$ has only one column), while the result can be straightforwardly extended to multi-input systems. A quantum system of (\ref{eq7}) and (\ref{eq8}) with the initial state $\mathbf{x}(0)=\mathbf{x}_0$ is equivalent to the following zero-initial-state system:
$$\left\{
\begin{array}{rl}
\mathbf{\dot{x}}&=A\mathbf{x}+Bu,\ \ \mathbf{x}(0)=\mathbf{0},\\
\mathbf{y}&=C\mathbf{x},\\
\end{array}
\right.$$
where $B=\mathbf{x}_0$ and ${u}=\delta(t)$.

For a quantum Hamiltonian, $\mathbf{x}_0$ and $C$ are usually determined and $A$ is antisymmetric. We rewrite (\ref{eq14}) as:
\begin{equation}\label{eq01}
SA(\mathbf\theta)=A(\mathbf{\theta'})S,
\end{equation}
\begin{equation}\label{eq02}
S\mathbf{x}_0=\mathbf{x}_0,
\end{equation}
\begin{equation}\label{eq03}
C=CS,
\end{equation}
together with the requirement that $S$ is nonsingular and other possible constraints on $\mathbf\theta$ and $\mathbf\theta'$. Eqs. (\ref{eq01})-(\ref{eq03}) are the starting point for STA analysis for the rest of this paper.

Next we use STA to test the identifiability for single-probe-assisted spin-1/2 chain systems in \cite{sone 2017}, which have the form of a one-dimensional chain, composed of multi qubits with their interaction governed by the system Hamiltonian. It is usually assumed that only the first qubit (the probe qubit) can be initialized and measured, while the rest qubits are all inaccessible (and thus they are assumed to be in the maximally mixed state initially). As in \cite{sone 2017}, we identify only the magnitude of the unknown parameters in the Hamiltonian; i.e., a system is identifiable if and only if all the solutions to the STA equations satisfy $|\theta_i|=|\theta_i'|$. There are four physical models in \cite{sone 2017}, where the transfer function on the Ising model without transverse field can be directly calculated and we omit the STA analysis for this model. The Ising model with the transverse field can also be skipped, because the system matrix has the same structure as that in the exchange model without transverse field. Hence, we only analyze two exchange models, with and without transverse field. Let $\mathbf{\theta}=(\theta_1,\theta_2,...,\theta_n)^T$ be the unknown parameters. For the exchange model without transverse field, $n+1$ is the total qubit number and the Hamiltonian can be written as
\begin{equation}\label{exchange1}
H=\sum_{i=1}^{n}\frac{(-1)^i\theta_i}{2}(X_iX_{i+1}+Y_{i}Y_{i+1}),
\end{equation}
where the subscript $i$ denotes the $i$th qubit, $X$ and $Y$ are the single-qubit Pauli matrices $$X=\left(\begin{matrix}
0&1\\
1&0\\
\end{matrix}\right),\ \ Y=\left(\begin{matrix}
0&-i\\
i&0\\
\end{matrix}\right).$$ The observable is $X_1$ with the initial state being an eigenstate of $X_1$. For the exchange model with transverse field, $n$ must be odd and $\frac{n+1}{2}$ is the total qubit number. The Hamiltonian can be written as
\begin{equation}\label{exchange2}
H=\sum_{i=1}^{\frac{n+1}{2}}\frac{\theta_{2i-1}}{2}Z_{i} +\sum_{i=1}^{\frac{n-1}{2}}\frac{\theta_{2i}}{2}(X_{i}X_{i+1}+Y_{i}Y_{i+1}),
\end{equation}
where $Z=\left(\begin{matrix}
1&0\\
0&-1\\
\end{matrix}\right)$. With the initial state being the eigenstate of $X_1$, the observable can be $X_1$ or $Y_1$. Therefore there are altogether three situations to be analyzed, which are summarized as Theorems \ref{th1}-\ref{th3}. These three situations were first investigated in \cite{sone 2017} and only verified numerically for several specific cases. Here, we provide a mathematical proof for arbitrary dimension. Also, Theorems \ref{th1}-\ref{th3} contain various situations to showcase the power of STA: Theorem \ref{th1} and Theorem \ref{th3} characterize identifiable minimal systems, while Theorem \ref{th2} corresponds to an unidentifiable minimal system. An example of dealing with identifiable non-minimal systems will be presented in Theorem \ref{th5}.

\section{EXCHANGE MODEL WITHOUT TRANSVERSE FIELD}\label{seccase1}
The Hamiltonian for this spin system is described in \cite{sone 2017}, which also derives the system model (\ref{exchange1}). Therefore we start from the linear system form (\ref{eq10}). In the system matrix $A$ only the elements directly above or below the main diagonal are non-zero:
\begin{equation}\label{eq18}
A=\left(\begin{array}{*{5}{c}}
0 & \theta_1 & 0 & 0 & \cdots \\
-\theta_1 & 0 & \theta_2 & 0 & \cdots\\
0 & -\theta_2 & 0 & \ddots & \\
0 & 0 & \ddots & & \theta_n \\
\vdots & \vdots &  &  -\theta_n & 0\\
\end{array}\right)_{(n+1)\times (n+1)}.
\end{equation}
The initial state of the probe is an eigenstate of $X_1$. Hence, $B=\mathbf{x}_0=(1,0,...,0)^T$. We measure $X_1$, and $C=(1,0,...,0)$. We have the following theorem:
\begin{theorem}\label{th1}
The exchange model without transverse field is identifiable when measuring $X_1$ on the single qubit probe.
\end{theorem}

\begin{IEEEproof}
We first prove this system is minimal for almost any value of the unknown parameters, and then test the identifiability.
\subsubsection{Proof for minimality}
\begin{lemma}\label{lem1}
With (\ref{eq18}) and $B=(1,0,...,0)^T$, the controllability matrix $\text{CM}=[B\ \ AB\ \ \cdots\ \ A^nB]$ has full rank for almost any value of $\mathbf\theta$.
\end{lemma}

The proof of Lemma \ref{lem1} is provided in Appendix \ref{plem1}. Then, given the observability matrix
\begin{equation}
\text{OM}=\left(
\begin{array}{c}
C\\
CA\\
\vdots\\
CA^n\\
\end{array}\nonumber\right)=\text{diag}(1,-1,1,-1,...,(-1)^n)\cdot \text{CM}^T,
\end{equation}
the system is also almost always observable. Therefore, it is almost always minimal.
\subsubsection{Identifiability test}
We now employ the STA equations to test the identifiability. Using (\ref{eq02}) and (\ref{eq03}) we know $S$ is of the form
\begin{equation}\label{eq19}
S=\left(
\begin{array}{cccc}
1 & 0 & \cdots & 0\\
0 & * & \cdots & *\\
\vdots & \vdots &  & \vdots\\
0 & * & \cdots & *\\
\end{array}
\right)_{(n+1)\times(n+1)},
\end{equation}
and (\ref{eq01}) is now
\begin{equation}\label{eq20}
\begin{array}{rl}
&\left(
\begin{array}{cccc}
1 & 0 & \cdots & 0\\
0 & * & \cdots & *\\
\vdots & \vdots &  & \vdots\\
0 & * & \cdots & *\\
\end{array}
\right)
\left(
\begin{array}{ccccc}
0 & \theta_1 & 0 & \cdots & 0\\
-\theta_1 & 0 & \ddots & &\\
0 & \ddots &  & &\\
\vdots &  &  & &\\
\end{array}
\right)\\
=&\left(
\begin{array}{ccccc}
0 & \theta_1' & 0 & \cdots & 0\\
-\theta_1' & 0 & \ddots & &\\
0 & \ddots &  & &\\
\vdots &  &  & &\\
\end{array}
\right)
\left(
\begin{array}{cccc}
1 & 0 & \cdots & 0\\
0 & * & \cdots & *\\
\vdots & \vdots &  & \vdots\\
0 & * & \cdots & *\\
\end{array}
\right).\\
\end{array}
\end{equation}
Denote the partitioned $S$ and $A$ as $$S=\left(\begin{matrix}1_{1\times 1}&(\mathbf{0}_{n\times 1})^T\\ \mathbf{0}_{n\times 1}&\tilde{S}_{n\times n}\end{matrix}\right),\ A=\left(\begin{matrix}0_{1\times 1}& (\mathbf{E}_{n\times 1})^T \\ -\mathbf{E}_{n\times 1} &\tilde{A}_{n\times n}\end{matrix}\right),$$
and then (\ref{eq20}) is equivalent to
\begin{equation}\label{eq21}
\mathbf{E}^T=\mathbf{E}'^T\tilde S,
\end{equation}
\begin{equation}\label{eq22}
-\tilde S \mathbf{E}=-\mathbf{E}',
\end{equation}
\begin{equation}\label{eq23}
\tilde S\tilde A=\tilde A'\tilde S.
\end{equation}
From the first elements in (\ref{eq21}) and (\ref{eq22}), we have $\theta_1=\theta_1'\tilde{S}_{11}$ and $-\tilde{S}_{11}\theta_1=-\theta_1'$. Since the atypical case of $\theta_1=0$ is not considered, we have $\theta_1'\neq0$ and $|\tilde{S}_{11}|=1$, which indicates $|\theta_1|=|\theta_1'|$. Then from the remaining elements in (\ref{eq21}) and (\ref{eq22}), we have $\tilde{S}_{12}=\tilde{S}_{13}=...=\tilde{S}_{1n}=0$ and $\tilde{S}_{21}=\tilde{S}_{31}=...=\tilde{S}_{n1}=0$.

If $\tilde{S}_{11}=1$, (\ref{eq23}) now is of the same form as (\ref{eq20}) but with dimension decreased by $1$; otherwise if $\tilde{S}_{11}=-1$, (\ref{eq23}) is equivalent to $(-\tilde S)\tilde A=\tilde A'(-\tilde S)$, which is also of the same form as (\ref{eq20}) with the dimension decreased by $1$. Therefore these procedures can be performed inductively and finally we know all the solutions to (\ref{eq20}) satisfy $S=\text{diag}(1,\pm1,...,\pm1)$ and $|\theta_i|=|\theta_i'|$ for all $1\leq i\leq n$.

\end{IEEEproof}

\begin{remark}
The relevant result in Theorem \ref{th1} was also presented in \cite{franco 2009}, where a specific Hamiltonian identification algorithm for the same system setting was proposed. Here we use it as an example to illustrate the effectiveness of STA.
\end{remark}

\section{EXCHANGE MODEL WITH TRANSVERSE FIELD}\label{seccase2}
The Hamiltonian for this system is as in (\ref{exchange2}) and we start from the linear system form (\ref{eq10}). In $A$, each $\theta_{2k+1}$ appears twice and each $\theta_{2k}$ appears four times:
\begin{equation}\label{eq71}
A=\left(\begin{array}{*{5}{c}}
0 & \theta_1 & 0 & -\theta_2 & \cdots \\
-\theta_1 & 0 & \theta_2 & 0 & \cdots\\
0 & -\theta_2 & 0 & \ddots & \\
\theta_2 & 0 & \ddots & & \theta_n \\
\vdots & \vdots &  &  -\theta_n & 0\\
\end{array}\right)_{(n+1)\times (n+1)},
\end{equation}
where $n$ must be odd. The initial state of the probe is an eigenstate of $X_1$. Hence, $B=\mathbf{x}_0=(1,0,...,0)^T$. With Property \ref{property1}, we can first rearrange $A$ as follows: we take its odd rows in ascending sequence and then take its even rows in ascending sequence, and we apply the same procedures to its columns. We rewrite $A$ into
\begin{equation}\label{eqad72}
A=\left(
\begin{matrix}
0 & \bar{A}\\
-\bar{A} & 0\\
\end{matrix}
\right),
\end{equation}
where
\begin{equation}\label{eq72}
\bar{A}=\left(
\begin{matrix}
\theta_1 & -\theta_2 & 0 &\cdots &0\\
-\theta_2 & \theta_3 & -\theta_4 & &\vdots\\
0 & -\theta_4 & \theta_5 & \ddots& 0\\
\vdots & & \ddots & \ddots &-\theta_{n-1}\\
 0 &\cdots  & 0& -\theta_{n-1}&\theta_n\\
\end{matrix}
\right)
\end{equation}
is symmetric. After this transformation, we have $B=(1,0,...,0)^T$ unchanged.

\subsection{Measuring $X_1$}
First we consider measuring $X_1$. Then $C=(1,0,...,0)$. We have the following conclusion:
\begin{theorem}\label{th2}
The exchange model with transverse field is unidentifiable when measuring $X_1$ on the single qubit probe.
\end{theorem}

\begin{IEEEproof}
We employ Criterion \ref{cri2} to prove the conclusion, and thus do not need to analyze its minimality. When $A$ in (\ref{eq71}) is transformed to (\ref{eqad72}), $C$ is unchanged and we assume $S$ is transformed to $\bar{S}$. Now (\ref{eq02}) and (\ref{eq03}) imply $\bar{S}$ is of the same form as (\ref{eq19}). We do not need to find all the solutions to (\ref{eq01}). Instead, we only need to find a special solution to (\ref{eq01}) which gives $|\theta_i|\neq|\theta_i'|$ for some $i$. We assume $$\bar{S}=\text{diag}({1_{1\times1},N_{\frac{n-1}{2}\times\frac{n-1}{2}},M_{\frac{n+1}{2}\times\frac{n+1}{2}}}),$$ which satisfies the form (\ref{eq19}). Eq. (\ref{eq01}) now is
\begin{equation}\label{eq73}
\left(
\begin{matrix}
1 & &\\
& N & \\
& & M\\
\end{matrix}
\right)\left(
\begin{matrix}
& \bar{A}\\
-\bar{A} & \\
\end{matrix}
\right)=\left(
\begin{matrix}
& \bar{A}'\\
-\bar{A}' & \\
\end{matrix}
\right)\left(
\begin{matrix}
1 & &\\
& N & \\
& & M\\
\end{matrix}
\right).
\end{equation}
We further assume $N$ and $M$ are orthogonal, which guarantees that $\bar{S}$ is nonsingular and now (\ref{eq73}) is in essence only one equation:
\begin{equation}\label{eq74}
\left(\begin{matrix}
1 & \\
& N\\
\end{matrix}\right)\bar{A}M^T=\bar{A}'.
\end{equation}

We perform spectral decomposition on $\bar{A}$ to have $\bar{A}=PEP^T$ where $P$ is orthogonal and $E$ is diagonal. We have the following lemma (the proof is given in Appendix \ref{appe}) to exclude the atypical cases:
\begin{lemma}\label{lem16}
Given arbitrary $\lambda_0\in\mathbb{C}$, it is atypical that $\lambda_0\in\Lambda(\bar{A})$.
\end{lemma}

Lemma \ref{lem16} is non-trivial. For example, if we change the structure of $\bar{A}$ as $\left(\begin{matrix}
\theta_1 & \theta_2\\
\theta_1 & \theta_2\\
\end{matrix}\right)$, then it is always true that $0\in\Lambda(\bar{A})$.

Denote $\mathbb{I}_k=\text{diag}(1,...,1,-1,1,...,1)$ where only the $k$th element is $-1$. We have the following assertion:
\begin{lemma}\label{lem17}
There is at least one $k\in \{1,2,...,n\}$ such that $|\theta_1|\neq|(PE\mathbb{I}_kP^T)_{11}|$.
\end{lemma}

The proof of Lemma \ref{lem17} is given in Appendix \ref{plem17}. Using Lemma \ref{lem17}, suppose $|\theta_1|\neq|(PE\mathbb{I}_mP^T)_{11}|$. We let $$M^T=P\mathbb{I}_mP^T\left(\begin{matrix}
1 & \\
& N^T\\
\end{matrix}\right).$$ As long as $N$ is orthogonal, $M$ is orthogonal. We denote the LHS of (\ref{eq74}) as $\bar{L}$, and have
\begin{equation}\label{eq75}
\begin{array}{rl}
\bar{L}&=\left(\begin{matrix}
1 & \\
& N\\
\end{matrix}\right)\bar{A}M^T\\
&=\left(\begin{matrix}
1 & \\
& N\\
\end{matrix}\right)PEP^TP\mathbb{I}_mP^T\left(\begin{matrix}
1 & \\
& N^T\\
\end{matrix}\right)\\
&=\left(\begin{matrix}
1 & \\
& N\\
\end{matrix}\right)PE\mathbb{I}_mP^T\left(\begin{matrix}
1 & \\
& N^T\\
\end{matrix}\right).\\
\end{array}
\end{equation}
We thus know
\begin{equation}
\begin{array}{rl}
|\bar{L}_{11}|&=\left|I_{1\sigma}\left(\begin{matrix}
1 & \\
& N\\
\end{matrix}\right)PE\mathbb{I}_mP^T\left(\begin{matrix}
1 & \\
& N^T\\
\end{matrix}\right)I_{\sigma1}\right|\\
&=|I_{1\sigma}PE\mathbb{I}_mP^TI_{\sigma1}|=|(PE\mathbb{I}_mP^T)_{11}|\neq|\theta_1|.\\
\end{array}\nonumber
\end{equation}
From (\ref{eq75}) we know $\bar{L}$ is always symmetric. Then we only need to find an appropriate orthogonal $N$ to make $\bar{L}$ have the same positions of zeros as $\bar{A}$. Denote $Z=PE\mathbb{I}_mP^T$, which is symmetric. We design a series of orthogonal matrices $N_{\frac{n-1}{2}\times\frac{n-1}{2}}^{(1)},N_{\frac{n-3}{2}\times\frac{n-3}{2}}^{(2)},...,N_{2\times2}^{(\frac{n-3}{2})}$ such that
$$N=\left(\begin{matrix}
I_{\frac{n-5}{2}\times\frac{n-5}{2}} &\\
 & N^{(\frac{n-3}{2})}\\
\end{matrix}\right)\cdots\left(\begin{matrix}
 I_{1\times1}& \\
 &N^{(2)}\\
\end{matrix}\right)N^{(1)}.$$
We further denote a series of ${\frac{n+1}{2}}$-dimensional matrices $Z^{(1)},Z^{(2)},...,Z^{(\frac{n-3}{2})}$ such that \begin{equation}\label{eq77}
Z^{(1)}=\left(\begin{matrix}
 1& \\
& N^{(1)}\\
\end{matrix}\right)Z\left(\begin{matrix}
 1& \\
& [N^{(1)}]^T\\
\end{matrix}\right)
\end{equation}
and $Z^{(i+1)}=\text{diag}(I_{i+1},N^{(i+1)})Z^{(i)}\text{diag}(I_{i+1},[N^{(i+1)}]^T)$ for $1\leq i\leq \frac{n-5}{2}$. Then $Z^{(\frac{n-3}{2})}=\bar{L}$. We start from the innermost layer (\ref{eq77}).

We partition $Z$ as $$Z=\left(\begin{matrix}
 Z_{11}& J_{1\times\frac{n-1}{2}}\\
(J_{1\times\frac{n-1}{2}})^T & \mathcal{J}_{\frac{n-1}{2}\times\frac{n-1}{2}}\\
\end{matrix}\right),$$ and have
\begin{equation}\label{eq78}
Z^{(1)}=\left(\begin{matrix}
 Z_{11}&J[N^{(1)}]^T \\
N^{(1)}J^T& N^{(1)}\mathcal{J}[N^{(1)}]^{T}\\
\end{matrix}\right).
\end{equation}
In (\ref{eq78}), $Z_{11}$ is unchanged and we need to make $J[N^{(1)}]^{T}$ have the form
\begin{equation}\label{eq79}
J[N^{(1)}]^{T}=(*,0,...,0).
\end{equation}
We perform spectral decomposition to set $$J^TJ=U^{(1)}\text{diag}(*,0,...,0)[U^{(1)}]^{T}.$$ Then $N^{(1)}=[U^{(1)}]^{T}$ is orthogonal and (\ref{eq79}) holds.

For the next layer, we partition $Z^{(1)}$ as
$$Z^{(1)}=\left(\begin{matrix}
Z_{11} & * & 0_{1\times\frac{n-3}{2}}\\
* & * & K_{1\times\frac{n-3}{2}}\\
0_{\frac{n-3}{2}\times1} & (K_{1\times\frac{n-3}{2}})^T & \mathcal{K}_{\frac{n-3}{2}\times\frac{n-3}{2}}\\
\end{matrix}\right).$$
We then have
\begin{equation}
\begin{array}{rl}
Z^{(2)}&=\left(\begin{matrix}
 1& &\\
 &1&\\
& &N^{(2)}\\
\end{matrix}\right)Z^{(1)}\left(\begin{matrix}
 1& &\\
 &1&\\
&&[N^{(2)}]^{T}\\
\end{matrix}\right)\\
&=\left(\begin{matrix}
Z_{11} & * & 0_{1\times\frac{n-3}{2}}\\
* & * & K[N^{(2)}]^{T}\\
0_{\frac{n-3}{2}\times1} & N^{(2)}K^T & N^{(2)}\mathcal{K}[N^{(2)}]^{T}\\
\end{matrix}\right).\\
\end{array}\nonumber
\end{equation}
$Z_{11}$ is unchanged and we need to make $K[N^{(2)}]^{T}$ take the form
$$K[N^{(2)}]^{T}=(*,0,...,0).$$
We perform spectral decomposition to make $$K^TK=U^{(2)}\text{diag}(*,0,...,0)[U^{(2)}]^{T},$$ and then $N^{(2)}=[U^{(2)}]^{T}$ is what we need. Continuing the above procedure, we can finally determine an orthogonal $N$ such that $\bar{L}=Z^{(\frac{n-3}{2})}$ has the same structure as $\bar{A}$. Since $Z_{11}$ is unchanged and $|Z_{11}|\neq|\theta_1|$, we know $|\bar{L}_{11}|\neq|\theta_1|$, which implies we have found a special unequal solution to the STA equations. Thus the system is unidentifiable.

\end{IEEEproof}

\subsection{Measuring $Y_1$}
Now we consider measuring $Y_1$, which sets $C=(0,1,0,...,0)$. We have the following theorem to correct the conclusion in \cite{sone 2017}.
\begin{theorem}\label{th3}
The exchange model with transverse field is identifiable when measuring $Y_1$ on the single qubit probe.
\end{theorem}

\begin{IEEEproof}
\subsubsection{Proof for minimality}
After $A$ in (\ref{eq71}) is transformed to (\ref{eqad72}), $C$ is transformed to
\begin{equation}\label{eq84}
{C}=(0_{1\times\frac{n+1}{2}},\bar{C}),\ \ \bar{C}=(1,0_{1\times\frac{n-1}{2}}).
\end{equation}
Denote
\begin{equation}\label{eq85}
B=(\bar{B}^T,0_{1\times\frac{n+1}{2}})^T,\ \ \bar{B}=(1,0_{1\times\frac{n-1}{2}})^T.
\end{equation}
We have the following lemma (the proof is given in Appendix \ref{appg}) to show that the system is minimal.

\begin{lemma}\label{lem19}
With (\ref{eqad72}), (\ref{eq72}), (\ref{eq84}) and (\ref{eq85}), both the controllability matrix $\text{CM}=[B\ \ AB\ \ \cdots\ \ A^nB]$ and the observability matrix $\text{OM}=[C^T\ \ A^TC^T\ \ \cdots\ \ A^{nT}C^T]^T$ have full rank for almost any value of $\mathbf{\theta}$.
\end{lemma}

\subsubsection{Identifiability test}
By Property \ref{property1}, we use STA to prove the system (\ref{eqad72}) and (\ref{eq72}) is identifiable with (\ref{eq84}) and (\ref{eq85}). We partition $S$ as
$$S=\left(\begin{matrix}
X_{\frac{n+1}{2}\times\frac{n+1}{2}} & *_{\frac{n+1}{2}\times\frac{n+1}{2}}\\
*_{\frac{n+1}{2}\times\frac{n+1}{2}} & Y_{\frac{n+1}{2}\times\frac{n+1}{2}}\\
\end{matrix}\right).$$
Then (\ref{eq01}) is
\begin{equation}\label{eq87}
\left(\begin{matrix}
X & *\\
* & Y\\
\end{matrix}\right)\left(\begin{matrix}
0 & \bar{A}\\
-\bar{A} & 0\\
\end{matrix}\right)=\left(\begin{matrix}
0 & \bar{A}'\\
-\bar{A}' & 0\\
\end{matrix}\right)\left(\begin{matrix}
X & *\\
* & Y\\
\end{matrix}\right),
\end{equation}
which is
\begin{equation}\label{eq88}
X\bar{A}=\bar{A}'Y,
\end{equation}
\begin{equation}\label{eq89}
Y\bar{A}=\bar{A}'X,
\end{equation}
where the other two equations on the indeterminate submatrices are omitted. Using (\ref{eq02}) and (\ref{eq03}), we have
\begin{equation}\label{eq90}
X_{\sigma1}=(1,0,...,0)^T,\ \ Y_{1\sigma}=(1,0,...,0).
\end{equation}

From (\ref{eq88}) and (\ref{eq89}), we have
\begin{equation}\label{eq91}
X^TX\bar{A}=X^T\bar{A}'Y=\bar{A}Y^TY,
\end{equation}
\begin{equation}\label{eq92}
Y^TY\bar{A}=Y^T\bar{A}'X=\bar{A}X^TX.
\end{equation}
From (\ref{eq91}) and (\ref{eq92}), the following relationship holds,
\begin{equation}\label{eq93}
(X^TX-Y^TY)\bar{A}=-\bar{A}(X^TX-Y^TY),
\end{equation}
which is a special form of Sylvester equation. We rephrase the general solving procedures for Sylvester equation \cite{zhou 1996} to solve (\ref{eq93}). We vectorize (\ref{eq93}) to have
$$(\bar{A}\otimes I_{\frac{n+1}{2}}+I_{\frac{n+1}{2}}\otimes\bar{A})\text{vec}({X^TX-Y^TY})=0.$$
Using the same idea in Appendices \ref{appe} and \ref{appg}, it is straightforward to prove that $\bar{A}\otimes I+I\otimes\bar{A}$ is almost always nonsingular by considering $\bar{A}=I$. An equivalent expression is that we almost always have
\begin{equation}\label{eq95}
\lambda_{i}(\bar{A})+\lambda_{j}(\bar{A})\neq0
\end{equation}
for any $1\leq i,j\leq \frac{n+1}{2}$. Therefore we can almost always have
\begin{equation}\label{eq96}
X^TX=Y^TY.
\end{equation}
Similarly,
$$(XX^T-YY^T)\bar{A}'=-\bar{A}'(XX^T-YY^T),$$
and thus
\begin{equation}\label{eq98}
(\bar{A}'\otimes I_{\frac{n+1}{2}}+I_{\frac{n+1}{2}}\otimes\bar{A}')\text{vec}({XX^T-YY^T})=0.
\end{equation}

\begin{lemma}\label{lem20}
With (\ref{eqad72}), (\ref{eq72}) and (\ref{eq87}), $\bar{A}'\otimes I_{\frac{n+1}{2}}+I_{\frac{n+1}{2}}\otimes\bar{A}'$ is almost always nonsingular.
\end{lemma}

The proof of Lemma \ref{lem20} is provided in Appendix \ref{apph}. With Lemma \ref{lem20}, we can almost always solve (\ref{eq98}) to have
\begin{equation}\label{eq99}
XX^T=YY^T.
\end{equation}
Considering (\ref{eq90}), we partition $X$ and $Y$ as
$$X=\left(\begin{matrix}
1_{1\times1} & E_{1\times\frac{n-1}{2}}\\
0_{\frac{n-1}{2}\times1} & \tilde{X}_{\frac{n-1}{2}\times\frac{n-1}{2}}\\
\end{matrix}\right),\ \ Y=\left(\begin{matrix}
1_{1\times1} & 0_{1\times\frac{n-1}{2}}\\
F_{\frac{n-1}{2}\times1} & \tilde{Y}_{\frac{n-1}{2}\times\frac{n-1}{2}}\\
\end{matrix}\right).$$
From (\ref{eq96}), $(X^TX)_{11}=1=(Y^TY)_{11}=1+F^TF$, which means $F=0$. Similarly from (\ref{eq99}) we have $E=0$. We partition $\bar{A}$ as
$$\bar{A}=\left(\begin{matrix}
\theta_1 & G_{1\times\frac{n-1}{2}}\\
(G_{1\times\frac{n-1}{2}})^T & \tilde{A}_{\frac{n-1}{2}\times\frac{n-1}{2}}\\
\end{matrix}\right).$$
Then (\ref{eq88}) is
$$\left(\begin{matrix}
1 & 0\\
0 & \tilde{X}\\
\end{matrix}\right)\left(\begin{matrix}
\theta_1 & G\\
G^T & \tilde{A}\\
\end{matrix}\right)=\left(\begin{matrix}
\theta_1' & G'\\
G'^T & \tilde{A}'\\
\end{matrix}\right)\left(\begin{matrix}
1 & 0\\
0 & \tilde{Y}\\
\end{matrix}\right),$$
which implies $\theta_1=\theta_1'$,
\begin{equation}\label{eq103}
G=G'\tilde{Y},
\end{equation}
\begin{equation}\label{eq104}
\tilde{X}G^T=G'^T,
\end{equation}
\begin{equation}\label{eq105}
\tilde{X}\tilde{A}=\tilde{A}'\tilde{Y}.
\end{equation}
Eq. (\ref{eq103}) is $(-\theta_2,0,...,0)=(-\theta_2',0,...,0)\tilde{Y}$, which implies $\tilde{Y}_{1\sigma}=(\theta_2/\theta_2',0,...,0)$. Similarly (\ref{eq104}) gives $\tilde{X}_{\sigma1}=(\theta_2'/\theta_2,0,...,0)^T$. With similar procedures, (\ref{eq89}) gives $\tilde{X}_{1\sigma}=(\theta_2/\theta_2',0,...,0)$, $\tilde{Y}_{\sigma1}=(\theta_2'/\theta_2,0,...,0)^T$ and
\begin{equation}\label{eq106}
\tilde{Y}\tilde{A}=\tilde{A}'\tilde{X}.
\end{equation}
Equating $\tilde{X}_{11}$ (or $\tilde{Y}_{11}$) we find $|\theta_2|=|\theta_2'|$. If $\theta_2=\theta_2'$, we have
\begin{equation}\label{eq107}
\tilde{Y}_{1\sigma}=(1,0,...,0)=(\tilde{X}_{\sigma1})^T.
\end{equation}
Now (\ref{eq105}), (\ref{eq106}) and (\ref{eq107}) have the same structures as (\ref{eq88}), (\ref{eq89}) and (\ref{eq90}), respectively, while with the dimension decreased by $1$. If $\theta_2=-\theta_2'$, we have
$-\tilde{Y}_{1\sigma}=(1,0,...,0)=(-\tilde{X}_{\sigma1})^T$ and we can rewrite (\ref{eq105}) and (\ref{eq106}) as
$(-\tilde{X})\tilde{A}=\tilde{A}'(-\tilde{Y})$ and $(-\tilde{Y})\tilde{A}=\tilde{A}'(-\tilde{X})$. Therefore, either $\{\tilde{X},\tilde{Y},\tilde{A},\tilde{A}'\}$ or $\{-\tilde{X},-\tilde{Y},\tilde{A},\tilde{A}'\}$ have the same structure and property as $\{X,Y,\bar{A},\bar{A}'\}$, but with the dimension decreased by $1$. This procedure can thus be performed recursively, until we finally reach $X=Y=\text{diag}(1,\pm1,...,\pm1)$ and $|\theta_i|=|\theta_i'|$ for every $1\leq i\leq n$.

\end{IEEEproof}
\begin{remark}
Theorem \ref{th2} and Theorem \ref{th3} indicate that when the system matrix $A$ has periodically repeated structure properties, STA analysis can avoid the curse of dimensionality and provide identifiability results for arbitrary dimension.
\end{remark}

\section{ECONOMIC QUANTUM HAMILTONIAN IDENTIFICATION ALGORITHMS}\label{secevi}
If a system is identifiable, we may develop an appropriate identification algorithm to identify the parameters. In this section, we provide another application of STA and SPT to quantum Hamiltonian identification. Generally the dimension of a quantum system is exponential in the number of qubits. Hence, identification algorithms that have polynomial complexity in the system dimension will in essence have exponential computational complexity in the number of qubits, which has been referred to as the exponential problem \cite{Nielsen and Chuang 2000}. To avoid this problem, one method is to design identification algorithms with computational complexity directly depending on quantities that increase much slower than the system dimension. Typically such quantities include the number of qubits in multi-qubit systems, or the number of unknown parameters for special physical systems. STA can be a useful tool to indicate the existence of such efficient algorithms.

\subsection{An indicator for the existence of efficient identification algorithms}\label{secevi1}

We aim to design an identification algorithm that has computational complexity that only depends on the number of unknown parameters. Suppose we have a $d$-dimensional Hamiltonian $H$ with ${\mathcal{M}}$ unknown parameters $\theta_i$. In most cases, the $a_i$s in (\ref{eq1}) are linear functions of $\theta_i$. Hence, we can expand $H$ directly using $\mathbf{\theta}$,
$$H=\sum_{i=1}^{\mathcal{M}} \theta_iH_i.$$
Using the procedures in Sec. \ref{subsec2}, we can model the evolution of the state as an $n$-dimensional linear system model
\begin{equation}\label{eq109}
\mathbf{\dot{x}}=A\mathbf{x},\ \ \mathbf{x}(0)=\mathbf{x_0},
\end{equation}
where each $\theta_i$ is an element of $A$. We hope the algorithm can identify one unknown element in $A$ under one set of $B$ and $C$, with computational complexity $f({\mathcal{M}})$ that is a function of ${\mathcal{M}}$ but not of $d$. Then the total computational complexity to identify the Hamiltonian is ${\mathcal{M}}f({\mathcal{M}})$, which does not directly depend on $d$.

We start by investigating the identification capability of the fundamental setting of $B=I_{\sigma i}$ and $C=I_{j\sigma}$. By changing indices, we assume that $B=I_{\sigma 2}$ and $C=I_{1\sigma}$. In the most general case, there are no special properties for the structure of $A$. Assume that this system $(A,B,C)$ is already minimal. Then from (\ref{eq02}) and (\ref{eq03}) we know the transformation matrix $S$ is
$$S=\left(\begin{matrix}
1&0&0&\cdots&0\\
*&1&*&\cdots&*\\
*&0&*&\cdots&*\\
\vdots&\vdots&\vdots&&\vdots\\
*&0&*&\cdots&*\\
\end{matrix}\right),$$
and (\ref{eq01}) is now
\begin{equation}\label{eqad109}
\begin{array}{rl}
&\left(\begin{matrix}
1&0&0&\cdots&0\\
*&1&*&\cdots&*\\
*&0&*&\cdots&*\\
\vdots&\vdots&\vdots&&\vdots\\
*&0&*&\cdots&*\\
\end{matrix}\right)\left(\begin{matrix}
A_{11}&A_{12}&\cdots\\
A_{21}&A_{22}&\cdots\\
\vdots&\vdots& \\
\end{matrix}\right)\\
&=\left(\begin{matrix}
A_{11}'&A_{12}'&\cdots\\
A_{21}'&A_{22}'&\cdots\\
\vdots&\vdots& \\
\end{matrix}\right)\left(\begin{matrix}
1&0&0&\cdots&0\\
*&1&*&\cdots&*\\
*&0&*&\cdots&*\\
\vdots&\vdots&\vdots&&\vdots\\
*&0&*&\cdots&*\\
\end{matrix}\right).\\
\end{array}
\end{equation}
By equating the elements on the first row and second column of both sides of (\ref{eqad109}), we have $A_{12}=A_{12}'$, which indicates this fundamental setting of $B$ and $C$ has the capability of identifying one parameter for minimal systems. Interestingly, we succeed in extending this conclusion to non-minimal systems using STA.

\begin{theorem}\label{th5}
Given a linear system $(A,B,C)$, $A_{ij}$ is identifiable (including its sign) if $B=I_{\sigma j}$ and $C=I_{i\sigma}$.
\end{theorem}

\begin{IEEEproof}
Without loss of generality, we can always assume that we are identifying $A_{12}$ or $A_{11}$ after appropriately changing the element order of $\mathbf{x}$ .

For the case of identifying $A_{12}$, $C=(1,0,...,0)$ and $B=(0,1,0,...,0)^T$. Without loss of generality, we assume that the system is neither controllable nor observable. We tentatively calculate the first two rows of the observability matrix, which are
\begin{equation}\label{eq110}
\left(\begin{matrix}
1&0&0&\cdots&0\\
*&A_{12}&*&\cdots&*\\
\end{matrix}\right).
\end{equation}
Since $A_{12}=0$ is atypical, it is almost always true that (\ref{eq110}) has rank two. Assume that the observable subsystem of (\ref{eq109}) has dimension $m$. We thus have $2\leq m<n$.

Let
$$T=\left(\begin{matrix}
1 & &&&&\\
& 1 &&&&\\
-A_{32}/A_{12} & &1&&&\\
-A_{42}/A_{12} & &&1&&\\
\vdots & &&&\ddots&\\
-A_{n2}/A_{12} & &&&&1\\
\end{matrix}\right)_{n\times n},$$
and perform a similarity transformation $\mathbf{\bar{x}}=T\mathbf{x}$. Using Property \ref{property1}, the equivalent system is
$$\bar{A}=TAT^{-1}=\left(\begin{matrix}
*&A_{12}&*&\cdots&*\\
*&*&*&\cdots&*\\
*&0&*&\cdots&*\\
\vdots&\vdots&\vdots&&\vdots\\
*&0&*&\cdots&*\\
\end{matrix}\right),$$
$\bar{B}=TB=(0,1,0,...,0)^T$ and $\bar{C}=CT^{-1}=(1,0,...,0)$. The former two rows in the observability matrix $\overline{\text{OM}}$ of the new system $(\bar{A},\bar{B},\bar{C})$ have the same form as (\ref{eq110}). Since $\overline{\text{OM}}$ has rank $m$, there exists a reordering $(j_3,j_4,...,j_n)$ of $(3,4,...,n)$ such that the matrix $(\overline{\text{OM}}_{\sigma1},\overline{\text{OM}}_{\sigma2},\overline{\text{OM}}_{\sigma j_3},\overline{\text{OM}}_{\sigma j_4},...,\overline{\text{OM}}_{\sigma j_m})$ is column-full-ranked. Let the matrix $U=(I_{\sigma1},I_{\sigma2},I_{\sigma j_3},I_{\sigma j_4},...,I_{\sigma j_n})^{-1}$ and perform a further similarity transformation $\mathbf{\tilde{x}}=U\mathbf{\bar{x}}$. Then the equivalent system is
\begin{equation}\label{eqad106}
\tilde{A}=U\bar{A}U^{-1}=\left(\begin{matrix}
*&A_{12}&*&\cdots&*\\
*&*&*&\cdots&*\\
*&0&*&\cdots&*\\
\vdots&\vdots&\vdots&&\vdots\\
*&0&*&\cdots&*\\
\end{matrix}\right)_{n\times n},
\end{equation}
$\tilde{B}=U\bar{B}=(0,1,0,...,0)^T$ and $\tilde{C}=\bar{C}U^{-1}=(1,0,...,0)$. Now the observability matrix of the system $\tilde{\Sigma}=(\tilde{A},\tilde{B}, \tilde{C})$ is
$$\widetilde{\text{OM}}=\left(\begin{matrix}
\tilde{C}\\
\tilde{C}\tilde{A}\\
\vdots\\
\tilde{C}\tilde{A}^{n-1}\\
\end{matrix}\right)=\left(\begin{matrix}
\bar{C}U^{-1}\\
\bar{C}\bar{A}U^{-1}\\
\vdots\\
\bar{C}\bar{A}^{n-1}U^{-1}\\
\end{matrix}\right)=\overline{\text{OM}}\cdot U^{-1}.$$
Therefore, the first $m$ columns of $\widetilde{\text{OM}}$ are of full-rank. We can now employ the SPT method. To perform observability decomposition for the system $\tilde{\Sigma}$, firstly we select the first two rows and other $m-2$ rows from $\widetilde{\text{OM}}$ to form a full-row-rank matrix $\tilde{E}_{m\times n}$ such that the former $m$ columns of $\tilde{E}$ are also full-rank. We partition $\tilde{E}$ as $\tilde{E}=[\tilde{F}_{m\times m}\ \ \mathbf{f}_{m\times(n-m)}]$, and then $\tilde{F}$ is invertible. The transformation matrix $\left(\begin{matrix}
\tilde{F}&\mathbf{f}\\
\mathbf{0}&I\\
\end{matrix}\right)$ can decompose the system $\tilde{\Sigma}$ into observable and unobservable parts. We choose the second transformation matrix as $$\text{diag}(\tilde{F}^{-1},I).$$ The total transformation is $$Q=\left(\begin{matrix} \tilde{F}^{-1} & \mathbf{0}\\ \mathbf{0}^T & I  \end{matrix}\right)\left(\begin{matrix} \tilde{F} & \mathbf{f}\\ \mathbf{0}^T & I  \end{matrix}\right)=\left(\begin{matrix} I& \tilde{F}^{-1}\mathbf{f} \\ \mathbf{0}^T & I  \end{matrix}\right),$$ and its inversion is $$Q^{-1}=\left(\begin{matrix} I& -\tilde{F}^{-1}\mathbf{f} \\ \mathbf{0}^T & I  \end{matrix}\right).$$ Let $\mathbf{\acute{x}}=Q\mathbf{\tilde{x}}$ generate the system $\acute{\Sigma}=(\acute{A},\acute{B},\acute{C})$:
$$\left\{
\begin{array}{rl}
\mathbf{\dot{\acute{x}}}&=\acute{A}\mathbf{\acute{x}}+\acute{B}{\delta(t)},\ \ \mathbf{\acute{x}}(0)=\mathbf{0},\\
y&=\acute{C}\mathbf{\bar{x}}.\\
\end{array}
\right.$$

We partition $\tilde{A}$ as
$$\tilde{A}=\left(\begin{matrix}
\widetilde{UL}_{m\times m} & \widetilde{UR}_{m\times (n-m)}\\
\widetilde{DL}_{(n-m)\times m} & \widetilde{DR}_{(n-m)\times (n-m)}\\
\end{matrix}\right).$$
Then we have
\begin{equation}
\begin{array}{rl}
\acute{A}&\!\!\!=Q\tilde{A}Q^{-1}=\left(\begin{matrix} I& \tilde{F}^{-1}\mathbf{f} \\ \mathbf{0}^T & I  \end{matrix}\right)\left(\begin{matrix}
\widetilde{UL} &\ \widetilde{UR}\\
\widetilde{DL} &\ \widetilde{DR}\\
\end{matrix}\right)\left(\begin{matrix} I& -\tilde{F}^{-1}\mathbf{f} \\ \mathbf{0}^T & I  \end{matrix}\right)\\
&\!\!\!=\left(\begin{matrix}
\widetilde{UL}+\tilde{F}^{-1}\mathbf{f}\widetilde{DL} &\ *_{m\times(n-m)}\\
*_{(n-m)\times m} &\ *_{(n-m)\times(n-m)}\\
\end{matrix}\right),\\
\end{array}\nonumber
\end{equation}
$\acute{B}=Q\tilde{B}=(0,1,0,...,0)^T$ and $\acute{C}=\tilde{C}Q^{-1}=(1,0,...,0,*,...,*)$, where elements in $\acute{C}$ from the second to the $m$th are all zero. We partition $\mathbf{\acute{x}}=(\mathbf{\grave{x}}^T,*)^T$ where $\mathbf{\grave{x}}$ is $m$-dimensional. Since the second transformation $\text{diag}(\tilde{F}^{-1},I)$ is block-diagonal, we know $\acute{\Sigma}$ is in the observable canonical form. Therefore, $\mathbf{\grave{x}}$ corresponds to the observable subsystem of $\acute{\Sigma}$. We denote this $m$-dimensional observable subsystem as $\grave{\Sigma}=(\grave{A},\grave{B},\grave{C})$ where $\grave{A}=\widetilde{UL}+\tilde{F}^{-1}\mathbf{f}\widetilde{DL}$, $\grave{B}=(0,1,0,...,0)^T$ and $\grave{C}=(1,0,...,0)$. From (\ref{eqad106}) we know $\widetilde{DL}_{\sigma2}=(0,0,...,0)^T$, and $\grave{A}_{\sigma 2}=\widetilde{UL}_{\sigma 2}$. Therefore, $\grave{A}_{12}=A_{12}$.

Similarly, we can employ the SPT method again to perform a controllability decomposition on $\grave{\Sigma}$ to finally obtain a $t$-dimensional ($2\leq t\leq m$) minimal system $(\check{A},\check{B},\check{C})$ where we still have $\check{A}_{12}=A_{12}$, $\check{B}=(0,1,0,...,0)^T$ and $\check{C}=(1,0,...,0)$.

For $(\check{A},\check{B},\check{C})$, we can employ the STA method. Using (\ref{eq02}) and (\ref{eq03}) we know the transformation matrix $S$ is
$$S=\left(\begin{matrix}
1&0&0&\cdots&0\\
*&1&*&\cdots&*\\
*&0&*&\cdots&*\\
\vdots&\vdots&\vdots&&\vdots\\
*&0&*&\cdots&*\\
\end{matrix}\right)_{t\times t},$$
and (\ref{eq01}) is now
\begin{equation}\label{eq119}
\begin{array}{rl}
&\left(\begin{matrix}
1&0&0&\cdots&0\\
*&1&*&\cdots&*\\
*&0&*&\cdots&*\\
\vdots&\vdots&\vdots&&\vdots\\
*&0&*&\cdots&*\\
\end{matrix}\right)\left(\begin{matrix}
*&A_{12}&*&\cdots&*\\
*&*&*&\cdots&*\\
\vdots&\vdots&\vdots&&\vdots\\
*&*&*&\cdots&*\\
\end{matrix}\right)\\
&=\left(\begin{matrix}
*&A_{12}'&*&\cdots&*\\
*&*&*&\cdots&*\\
\vdots&\vdots&\vdots&&\vdots\\
*&*&*&\cdots&*\\
\end{matrix}\right)\left(\begin{matrix}
1&0&0&\cdots&0\\
*&1&*&\cdots&*\\
*&0&*&\cdots&*\\
\vdots&\vdots&\vdots&&\vdots\\
*&0&*&\cdots&*\\
\end{matrix}\right)\\
\end{array}
\end{equation}
Equating the elements on the first row and second column of both sides of (\ref{eq119}), we have $A_{12}=A_{12}'$. Thus $A_{12}$ is identifiable.

For the case identifying $A_{11}$, $B^T=C=(1,0,...,0)$. Its observability matrix is now
$$\text{OM}=\left(\begin{matrix}
1&0&\cdots&0\\
A_{11}&*&\cdots&*\\
\hdotsfor{4} \\
\end{matrix}\right).$$
If $\text{OM}_{2\sigma}$ has non-zero elements other than $A_{11}$, then the former two rows of $\text{OM}$ are linearly independent and we can use similar procedures to the case of identifying $A_{12}$ to prove that $A_{11}$ is identifiable. Otherwise if $\text{OM}_{2\sigma}=(A_{11},0,...,0)$, then $A_{1\sigma}=(A_{11},0,...,0)$, which means $(A,B,C)$ now is already of the observable canonical form, where the observable subsystem is $1$-dimensional:
\begin{equation}
\left\{
\begin{array}{rl}
\dot{x}_1&=\theta_1{x_1}+1\cdot\delta(t),\ \ {x_1}(0)={0},\\
{y}&=1\cdot{x_1}.\\
\end{array}
\right.\nonumber
\end{equation}
Hence, $\theta_1$ is certainly identifiable, which completes the proof.

\end{IEEEproof}

\subsection{An economic Hamiltonian identification algorithm}

Theorem \ref{th5} indicates the existence of economic quantum Hamiltonian identification algorithms. A natural following question is whether we can develop an economic algorithm. In fact, the proof of Theorem \ref{th5} has already implied how to prepare the initial state of the system and select the observable. Here, we present an identification algorithm based on the Taylor expansion of matrix exponential function \cite{my 2017}.

We start from the system (\ref{eq109}) that has a solution $y(t)=Ce^{At}\mathbf{x_0}$. We assume that in actual experiments we can sample the system output with a fixed period of time $\Delta t$, and the data length is $\mathcal{N}$. Then the data we obtain is denoted as $D=(y(\Delta t), y(2\Delta t),...,y(\mathcal{N}\Delta t))^T$ and its $i$th element is $D_i=y(i\Delta t)$. To estimate $\theta_{i_0}=A_{jk}$, we prepare the system initial value in a state corresponding to $B=\mathbf{x_0}=I_{\sigma k}$ and measure the observable corresponding to $C=I_{j\sigma}$.

We rewrite the data as
\begin{equation}
\begin{array}{rl}
D_p&=Ce^{pA\Delta t}B=\sum_{r=0}^{\infty}\frac{p^r\Delta t^r}{r!}I_{j\sigma}A^rI_{\sigma k}\\
&=\delta_{jk}+\sum_{r=1}^{\infty}\frac{p^r\Delta t^r}{r!}(A^r)_{jk}\approx\delta_{jk}+\sum_{r=1}^{q}\frac{p^r\Delta t^r}{r!}(A^r)_{jk},\\
\end{array}\nonumber
\end{equation}
where we should choose $q\leq \mathcal{N}$.

Denote $w=\mathcal{N}||{\rm{A}}||\Delta te$ and $z=1+\max(\lfloor w\rfloor,q)$ for simplicity. We bound the truncated terms as
\begin{equation}
\begin{array}{rl}
&\ \ \ \ |\sum_{r=q+1}^{\infty}\frac{p^r{\Delta t}^r}{r!} ({\rm{A}}^r)_{ik}|\\
&\leq\sum_{r=q+1}^{\infty}|\frac{1}{\sqrt{2\pi r}}\frac{(p\Delta te)^r}{r^r}({\rm{A}}^r)_{ik}|\\
&=\sum_{r=q+1}^{\infty}\frac{1}{\sqrt{2\pi r}}(\frac{p\Delta te}{r})^r|I_{i\sigma}{\rm{A}}^rI_{\sigma k}|\\
&\leq\sum_{r=q+1}^{\infty}\frac{1}{\sqrt{2\pi r}}(\frac{p\Delta te}{r})^r||I_{i\sigma}||\cdot||{\rm{A}}||^r\cdot||I_{\sigma k}||\\
&=\sum_{r=q+1}^{\infty}\frac{1}{\sqrt{2\pi r}}(\frac{p||{\rm{A}}||\Delta te}{r})^r\\
&\leq\frac{1}{\sqrt{2\pi (q+1)}}\sum_{r=q+1}^{\infty}(\frac{w}{r})^r\\
&\leq\frac{1}{\sqrt{2\pi (q+1)}}\sum_{r=q+1}^{z-1}(\frac{w}{r})^r+\frac{1}{\sqrt{2\pi (q+1)}}\sum_{r=z}^{\infty}(\frac{w}{z})^r\\
&=\frac{1}{\sqrt{2\pi (q+1)}}\sum_{r=q+1}^{z-1}(\frac{w}{r})^r+\frac{(\frac{w}{z})^{z}}{\sqrt{2\pi (q+1)}(1-\frac{w}{z})},\\
\end{array}\nonumber
\end{equation}
where the first line comes from Stirling's approximation. Hence, the summation of the truncated items is never divergent.

Denote $\Psi^{(q)}=(\psi_1,\psi_2,...,\psi_q)^T$ where $\psi_i=(A^i)_{jk}$. Then we need to identify $\theta_{i_0}=A_{jk}=\psi_1$. Denote
$$L=\left(\begin{array}{*{4}{c}}
\frac{1^1{\Delta t}^1}{1!} & \frac{1^2{\Delta t}^2}{2!} & \cdots & \frac{1^q{\Delta t}^q}{q!} \\
\frac{2^1{\Delta t}^1}{1!} & \frac{2^2{\Delta t}^2}{2!} & \cdots & \frac{2^q{\Delta t}^q}{q!} \\
\multicolumn{4}{c}{\dotfill} \\
\frac{\mathcal{N}^1{\Delta t}^1}{1!} & \frac{\mathcal{N}^2{\Delta t}^2}{2!} & \cdots & \frac{\mathcal{N}^q{\Delta t}^q}{q!} \\
\end{array}\right)_{\mathcal{N}\times q}.$$
We have $D\approx L\Psi^{(q)}$. We use a least-squares method to obtain an estimate
$$\hat{\Psi}^{(q)}=(L^TL)^{-1}L^TD,$$
and $\hat{\theta}_{i_0}=\hat{\psi}_1$. To fully reconstruct any $H$, this algorithm has online computational complexity $O({\mathcal{M}}^2+q{\mathcal{M}}\mathcal{N})$. In the worst case, there is no prior knowledge on $H$ and the computational complexity becomes $O(d^4+d^2q\mathcal{N})$. As long as $\mathcal{N}=o(d^2)$, this computational complexity is lower than the $O(d^6)$ of the identification algorithm in \cite{my 2016}. For another example of such economic Hamiltonian identification algorithms, please refer to \cite{my ifac}.

\subsection{Numerical example}
\begin{figure}
\centering
\includegraphics[width=3.4in]{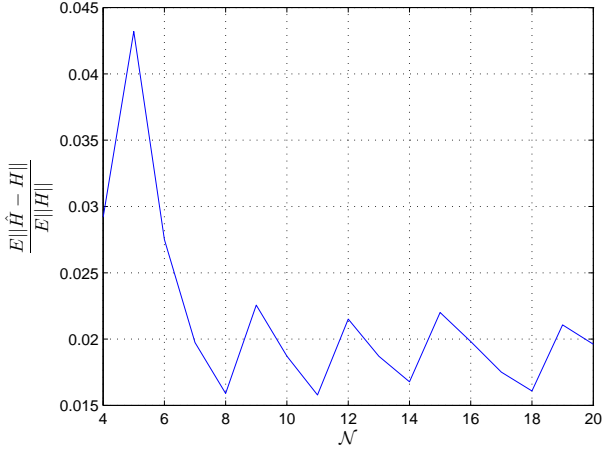}
\centering{\caption{Relative identification error $\frac{E||\hat H-H||}{E||H||}$ versus data length $\mathcal{N}$.}\label{test1}}
\end{figure}

We perform numerical simulations to illustrate the performance of the identification algorithm. Consider a 5-qubit exchange model without transverse field ($n=4$ in (\ref{exchange1})) and the values of the Hamiltonian parameters are $\mathbf{\theta}=(0.1,1.5,-0.8,3.1)$. The accessible set is $\bar G=\{X_1,Z_1Y_2,Z_1Z_2X_3,Z_1Z_2Z_3Y_4,Z_1Z_2Z_3Z_4X_5\}$. We set the initial states of the system as the eigenstates of $Z_1Y_2,Z_1Z_2X_3,Z_1Z_2Z_3Y_4,Z_1Z_2Z_3Z_4X_5$ and observe $X_1,Z_1Y_2,Z_1Z_2X_3,Z_1Z_2Z_3Y_4$, respectively. From Theorem \ref{th5} we know all the parameters are identifiable. Then we identify the Hamiltonian using the Taylor expansion identification algorithm. The sampling period is $\Delta t=0.1s$ and the parameter $q=\lfloor 0.3\mathcal{N}\rfloor+3$. We add zero-mean Gaussian noise with standard deviation $0.001$ into the sampling data. The identification result is shown in Fig. \ref{test1}, where each point is repeated $500$ times. In Fig. \ref{test1}, the horizontal axis is the data length $\mathcal{N}$ and the vertical axis is the relative identification error $\frac{E||\hat H-H||}{E||H||}$, where $E(\cdot)$ is the expectation on all the possible measurement results. The numerical result shows that the identification algorithm can effectively identify the Hamiltonian.

\section{CONCLUSION}\label{secfinal}
We have extended the STA method in classical control theory to the domain of quantum Hamiltonian identification, and employed the STA method to prove the identifiability of spin-1/2 chain systems assisted by single-qubit probes \cite{sone 2017}. STA has been demonstrated to be a powerful tool to analyze the identifiability for quantum systems with arbitrary dimension, which is also helpful for further designing identification algorithms. STA can also serve as a useful method for physicists to investigate the information extraction capability of quantum subsystems (like the single qubit probe in \cite{sone 2017}). An SPT method was developed to efficiently test the identifiability for non-minimal systems. We further employed the SPT method to provide an indicator for the existence of economic quantum Hamiltonian identification algorithms. The SPT method is proved to be a strong supplement to STA. SPT can also be applicable to classical control systems, especially when the experimental settings are difficult to change. We proposed an example of economic quantum Hamiltonian identification algorithms and presented a numerical example to illustrate the performance of the identification algorithm.

Future work includes developing a general framework using STA to characterize the amount of identifiable information for an unidentifiable system. It will also be helpful to propose more sufficient or necessary conditions for a system to be identifiable. Furthermore, it is useful to develop other efficient Hamiltonian identification algorithms with good performance.

\appendices

\section{PROOF OF LEMMA \ref{lem1}}\label{plem1}
\begin{IEEEproof}
By induction we have $$A^kB=[(*,...,*,(-1)^{k}\prod_{i=1}^{k}\theta_i,0,...,0)^T]_{(n+1)\times1}$$ for $1\leq k\leq n$ where $*$ are polynomials in $\theta_i$ and the last $n-k$ elements are zero. Therefore, $\text{CM}$ is an upper triangular matrix and its determinant is $$\det(\text{CM})=\prod_{k=1}^n (-1)^{k}\prod_{i=1}^{k}\theta_i,$$ which is non-zero for almost any value of $\mathbf\theta$. Hence, $\text{CM}$ is almost always full-ranked.

\end{IEEEproof}

\section{PROOF OF LEMMA \ref{lem16}}\label{appe}
\begin{IEEEproof}
We consider $\det(\bar{A}-\lambda_0I)$, which must equal to one of the following three possibilities:
(a) A non-trivial polynomial in $\theta_i$s ($i=1,2,...,n$);
(b) A non-zero constant;
(c) The constant zero.
We let $\theta_2=\theta_4=...=\theta_{n-1}=0$ and $\theta_1=\theta_3=...=\theta_{n}=\lambda_0+1$. Then from (\ref{eq72}) we know $\det(\bar{A}-\lambda_0I)=\det(I)=1$. Therefore, (c) is excluded. No matter which of (a) and (b) is valid, $\det(\bar{A}-\lambda_0I)\neq0$ for almost any value of $\mathbf{\theta}$, which implies that it is atypical to assume $\lambda_0\in\Lambda(\bar{A})$.

\end{IEEEproof}

\section{PROOF OF LEMMA \ref{lem17}}\label{plem17}
\begin{IEEEproof}
Since $\bar{A}=PEP^T=\sum_{i=1}^nE_{ii}P_{\sigma i}(P^T)_{i\sigma}$, we have
\begin{equation}\label{eqf1}
\theta_1=I_{1\sigma}\bar{A}I_{\sigma1}=\sum_{i=1}^nE_{ii}I_{1\sigma}P_{\sigma i}(P^T)_{i\sigma}I_{\sigma1}=\sum_{i=1}^nE_{ii}P_{1i}^2.
\end{equation}
Since $\sum_{i=1}^nP_{1i}^2=1$, $P_{1\sigma}$ can not be all zero. Suppose there are $m$ non-zero elements in $P_{1\sigma}$ where $1\leq m\leq n$. If $m=1$, we suppose it is $P_{1t}\neq0$. Then $P_{1t}=\pm1$ and $P_{1i}=0$ for every $i\neq t$. Since $\sum_{j=1}^nP_{jt}^2=1$, $P_{jt}=0$ for every $j\neq1$. We calculate
\begin{equation}
\begin{array}{rl}
-\theta_2&=I_{1\sigma}\bar{A}I_{\sigma2}=\sum_{i=1}^nE_{ii}I_{1\sigma}P_{\sigma i}(P^T)_{i\sigma}I_{\sigma2}\\
&=\sum_{i=1}^nE_{ii}P_{1i}P_{2i}=E_{tt}P_{1t}P_{2t}=0,\\
\end{array}\nonumber
\end{equation}
which is atypical and can be ignored. Hence, it is almost always true that $m\geq2$. We assume that $P_{1i_j}\neq0$ for $i_j=i_1,i_2,...,i_m$ and otherwise $P_{1i}=0$.

We prove the conclusion of Lemma \ref{lem17} by contradiction. Suppose for every $1\leq k\leq n$, $|\theta_1|=|(PE\mathbb{I}_kP^T)_{11}|$. Since
\begin{equation}
\begin{array}{rl}
&\ \ (PE\mathbb{I}_kP^T)_{11}=I_{1\sigma}[PEP^T-PE(I-\mathbb{I}_k)P^T]I_{\sigma1}\\
&\!\!\!\!=I_{1\sigma}[\sum_{i=1}^nE_{ii}P_{\sigma i}(P^T)_{i\sigma}-2E_{kk}P_{\sigma k}(P^T)_{k\sigma}]I_{\sigma1}\\
&\!\!\!\!=\sum_{i=1}^nE_{ii}P_{1i}^2-2E_{kk}P_{1k}^2\\
&\!\!\!\!=\theta_1-2E_{kk}P_{1k}^2,\\
\end{array}\nonumber
\end{equation}
we always have
\begin{equation}\label{eqf4}
|\theta_1|=|\theta_1-2E_{kk}P_{1k}^2|.
\end{equation}
We let $k=i_1$ in (\ref{eqf4}). From Lemma \ref{lem16}, we have $E_{i_1i_1}\neq0$. Since $P_{1i_1}\neq0$, we take the square of both sides of (\ref{eqf4}) and obtain $\theta_1=E_{i_1i_1}P_{1i_1}^2$. For the same reason, we have $\theta_1=E_{i_2i_2}P_{1i_2}^2=...=E_{i_mi_m}P_{1i_m}^2$. Then (\ref{eqf1}) implies $\theta_1=mE_{i_1i_1}P_{1i_1}^2$, which means $E_{i_1i_1}P_{1i_1}^2=0$ and implies a contradiction.

\end{IEEEproof}

\section{PROOF OF LEMMA \ref{lem19}}\label{appg}
\begin{IEEEproof}
The controllability matrix is
$$\text{CM}=\left(\begin{matrix}
\bar{B}&0&-\bar{A}^2\bar{B}&0&...& 0\\
0&-\bar{A}\bar{B}&0&\bar{A}^3\bar{B}&...& -\bar{A}(-\bar{A}^2)^{\frac{n-1}{2}}\bar{B}\\
\end{matrix}\right).$$
Hence, it suffices to prove that $Q=(\bar{B},\bar{A}^2\bar{B},...,\bar{A}^{n-1}\bar{B})$ is almost always nonsingular. Similar to the analysis in Appendix \ref{appe}, $\det(Q)$ has only three possibilities, where the possibility of $\det(Q)\equiv0$ needs to be excluded. Hence, we only need to find a special $\bar{A}$ such that $\det(Q)\neq0$.

We take
\begin{equation}
\bar{A}=\left(
\begin{array}{ccccc}
0&1&&&\\
1&0&1&&\\
&1&\ddots&\ddots&\\
&&\ddots&0&1\\
&&&1&1\\
\end{array}\right).\nonumber
\end{equation}
Then
\begin{equation}
\bar{A}^2=\left(
\begin{array}{ccccccc}
1&0&1&&&&\\
0&2&0&1&&&\\
1&0&2&0&\ddots&&\\
&1&&\ddots&&0&1\\
&&&\ddots&0&2&1\\
&&&&1&1&2\\
\end{array}\right).\nonumber
\end{equation}
We can take $Q$ as the controllability matrix of another system $(\bar{A}^2,\bar{B})$, which should be controllable. Since controllability is unchanged under similarity transformation, we transform $\bar{A}^2$ into
\begin{equation}\label{eqg4}
\tilde{A}=\left(
\begin{array}{ccccc}
1&1&&&\\
1&2&1&&\\
&1&\ddots&\ddots&\\
&&\ddots&2&1\\
&&&1&1\\
\end{array}\right).
\end{equation}
This similarity transformation works in the following steps: (i) We take all the odd rows of $\bar{A}^2$ in ascending order. (ii) Following (i), we take all the even rows of $\bar{A}^2$ in descending order. (iii) We repeat (i) and (ii) on the columns of $\bar{A}^2$. After steps (i) and (ii), each $2$ (except the $2$ in the last row) will have a $1$ just above it and a $1$ just below it, and this property does not change in step (iii). Also, the transformation is symmetric. Hence, $\tilde{A}$ is symmetric with all the $2$s on the diagonal line. $\tilde{A}$ thus has the form of (\ref{eqg4}). Under this transformation, $\tilde{B}=\bar{B}$ is unchanged.

For system $(\tilde{A},\tilde{B})$, it can be proven by induction that the controllability matrix $\tilde{Q}$ is an upper triangular matrix with all the diagonal elements $1$. Therefore $\det(\tilde{Q})\neq0$, and thus $\det(Q)\neq0$ and the possibility (c) is excluded. Hence, $\text{CM}$ is almost always full-rank.

For the observability matrix,
\begin{equation}
\text{OM}=\left(
\begin{array}{cc}
0&\bar{C}\\
-\bar{C}\bar{A}&0\\
0&-\bar{C}\bar{A}^2\\
\cdots&\cdots\\
-\bar{C}\bar{A}(-\bar{A}^2)^{\frac{n-1}{2}}&0\\
\end{array}\right).\nonumber
\end{equation}
Hence, it suffices to prove that $P=(\bar{C}^T,\bar{A}^{2T}\bar{C}^T,...,\bar{A}^{(n-1)T}\bar{C}^T)^T$ is almost always nonsingular. Since $\bar{A}$ is symmetric and $\bar{C}^T=\bar{B}$, we know $P=Q^T$. Therefore, $\text{OM}$ is also almost always full-rank.

\end{IEEEproof}

\section{PROOF OF LEMMA \ref{lem20}}\label{apph}
\begin{IEEEproof}
First, we investigate the relationship between $\Lambda(\bar{A})$ and $\Lambda(\bar{A})'$. Since $A$ is similar to $A'$, we know $A^2$ is similar to $A'^2$, which implies $\Lambda(A^2)=\Lambda(A'^2)$. Therefore, $$\Lambda\left(\begin{matrix}
-\bar{A}^2 &0\\
0& -\bar{A}^2\\
\end{matrix}\right)=\Lambda\left(\begin{matrix}
-\bar{A}'^2 &0\\
0& -\bar{A}'^2\\
\end{matrix}\right).$$ If we arrange the eigenvalues of $\bar{A}$ and $\bar{A}'$ both in ascending sequences, we have
\begin{equation}\label{eq201}
\lambda_i(\bar{A}')=p_i\lambda_i(\bar{A})
\end{equation}
for $1\leq i\leq \frac{n+1}{2}$ where $p_i=\pm1$.

Second, we point out that it is atypical for $\bar{A}$ to have multiple eigenvalues. We consider $\det(\lambda I-\bar{A})$, which is a polynomial on $\lambda$ with the coefficients being polynomials on $\theta_i$s. $\det(\lambda I-\bar{A})$ has multiple roots if and only if its discriminant, which is a polynomial function in the coefficients of $\det(\lambda I-\bar{A})$, equals zero \cite{gelfand 2008}. We can view this discriminant as a polynomial function in $\theta_i$s. If this discriminant is in fact the constant zero, then $\det(\lambda I-\bar{A})$ will always have multiple roots, which can be excluded by taking $\bar{A}=\text{diag}(1,2,...,\frac{n+1}{2})$. Therefore, the discriminant does not degenerate to zero, and its solution set is of zero measure. Hence, the set of $\mathbf{\theta}$ that can make $\det(\lambda I-\bar{A})$ have multiple roots is of zero measure, which implies that it is atypical when $\bar{A}$ has multiple eigenvalues.

Third, we prove that we can almost always have $\lambda_i(\bar{A}')+\lambda_j(\bar{A}')\neq0$ for any $1\leq i,j\leq\frac{n+1}{2}$. Using (\ref{eq201}) we have
\begin{equation}\label{eqh1}
\lambda_i(\bar{A}')+\lambda_j(\bar{A}')=p_i\lambda_i(\bar{A})+p_j\lambda_j(\bar{A}).\\
\end{equation}
If $i=j$, then the RHS of (\ref{eqh1}) is $2p_i\lambda_i(\bar{A})$, which is almost always non-zero according to Lemma \ref{lem16}. If $i\neq j$, the RHS of (\ref{eqh1}) is $p_i[\lambda_i(\bar{A})\pm\lambda_j(\bar{A})]$, which is also almost always non-zero because of (\ref{eq95}) and the fact that $\bar{A}$ almost always has no multiple eigenvalues. Therefore, we can almost always have $\lambda_i(\bar{A}')+\lambda_j(\bar{A}')\neq0$ for any $1\leq i,j\leq\frac{n+1}{2}$, which is equivalent to the statement that $\bar{A}'\otimes I_{\frac{n+1}{2}}+I_{\frac{n+1}{2}}\otimes\bar{A}'$ is almost always nonsingular.

\end{IEEEproof}

\section*{ACKNOWLEDGEMENT}

Y. Wang would like to thank Qi Yu for the helpful discussion.

\ifCLASSOPTIONcaptionsoff
  \newpage
\fi

\begin{thebibliography}{1}     

\bibitem{Nielsen and Chuang 2000}
M. A. Nielsen and I. L. Chuang, {\em Quantum Computation and Quantum Information.} Cambridge, U.K.: Cambridge Univ. Press, 2000.

\bibitem{quantum sensing}
C. L. Degen, F. Reinhard, and P. Cappellaro, ``Quantum sensing," {\em Rev. Mod. Phys.}, vol. 89, no. 3, 2017, Art. no. 035002.

\bibitem{paris 2004}
M. Paris and J. \v{R}eh\'{a}\v{c}ek, {\em Quantum State Estimation (Lecture Notes in Physics)}, vol. 649, Berlin, Germany: Springer, 2004.

\bibitem{Wiseman 2009}
H. M. Wiseman and G. J. Milburn, {\em Quantum Measurement and Control.} Cambridge, U.K.: Cambridge Univ. Press, 2009.

\bibitem{hou 2016}
Z. Hou, H.-S. Zhong, Y. Tian, D. Dong, B. Qi, L. Li, Y. Wang, F. Nori, G.-Y. Xiang, C.-F. Li, and G.-C. Guo, ``Full reconstruction of a 14-qubit state within four hours," {\em New J. Phys.}, vol. 18, 2016, Art. no. 083036.

\bibitem{qi 2017}
B. Qi, Z. Hou, Y. Wang, D. Dong, H.-S. Zhong, L. Li, G.-Y. Xiang, H. M. Wiseman, C.-F. Li, and G.-C. Guo, ``Adaptive quantum state tomography via linear regression estimation: Theory and two-qubit experiment," {\em npj Quantum Inform.}, vol. 3, no. 19, 2017.

\bibitem{sone 2018}
A. Sone, Q. Zhuang, and P. Cappellaro, ``Quantifying precision loss in local quantum thermometry via diagonal discord," {\em Phys. Rev. A}, vol. 98, no. 1, 2018, Art. no. 012115.

\bibitem{wang 2017}
J. Wang, S. Paesani, R. Santagati, S. Knauer, A. A. Gentile, N. Wiebe, M. Petruzzella, J. L. O'Brien, J. G. Rarity, A. Laing, and M. G. Thompson, ``Experimental quantum Hamiltonian learning," {\em Nat. Phys.}, vol. 13, no. 6, pp. 551-555, 2017.

\bibitem{devitt 2006}
S. J. Devitt, J. H. Cole, and L. C. Hollenberg, ``Scheme for direct measurement of a general two-qubit Hamiltonian," {\em Phys. Rev. A}, vol. 73, no. 5, 2006, Art. no. 052317.

\bibitem{zhang 2015}
J. Zhang and M. Sarovar, ``Identification of open quantum systems from observable time traces," {\em Phys. Rev. A}, vol. 91, no. 5, 2015, Art. no. 052121.

\bibitem{fu 2016}
Y. Fu, H. Rabitz, and G. Turinici, ``Hamiltonian identification in presence of large control field perturbations," {\em J. Phys. A: Math. Theor.}, vol. 49, no. 49, 2016, Art. no. 495301.

\bibitem{levitt 2018}
M. Levitt, M. Gu\c{t}\u{a}, and H. I. Nurdin, ``Power spectrum identification for quantum linear systems," {\em Automatica}, vol. 90, pp. 255-262, 2018.

\bibitem{bonnabel 2009}
S. Bonnabel, M. Mirrahimi, and P. Rouchon, ``Observer-based Hamiltonian identification for quantum systems," {\em Automatica}, vol. 45, no. 5, pp. 1144-1155, 2009.

\bibitem{shu 2016}
C.-C. Shu, K.-J. Yuan, D. Dong, I. R. Petersen, and A. D. Bandrauk, ``Identifying strong-field effects in indirect photofragmentation reactions," {\em J. Phys. Chem. Lett.}, vol. 8, no. 1, pp. 1-6, 2016.

\bibitem{burgarth 2012}
D. Burgarth and K. Yuasa, ``Quantum system identification," {\em Phys. Rev. Lett.}, vol. 108, no. 8, 2012, Art. no. 080502.

\bibitem{bris 2007}
C. Le Bris, M. Mirrahimi, H. Rabitz, and G. Turinici, ``Hamiltonian identification for quantum systems: Well-posedness and numerical approaches," {\em ESAIM: Control Optim. Calculus Variations}, vol. 13, no. 2, pp. 378-395, 2007.

\bibitem{franco 2011}
C. Di Franco, M. Paternostro, and M. S. Kim, ``Bypassing state initialization in Hamiltonian tomography on spin-chains," {\em Int. J. Quantum Inf.}, vol. 9, supp. 01, pp. 181-187, 2011.

\bibitem{franco 2009}
C. Di Franco, M. Paternostro, and M. S. Kim, ``Hamiltonian tomography in an access-limited setting without state initialization," {\em Phys. Rev. Lett.}, vol. 102, no. 18, 2009, Art. no. 187203.

\bibitem{burgarth 2009a}
D. Burgarth, K. Maruyama, and F. Nori, ``Coupling strength estimation for spin chains despite restricted access," {\em Phys. Rev. A}, vol. 79, no. 2, 2009, Art. no. 020305.

\bibitem{burgarth 2009}
D. Burgarth and K. Maruyama, ``Indirect Hamiltonian identification through a small gateway," {\em New J. Phys.}, vol. 11, no. 10, 2009, Art. no. 103019.

\bibitem{burgarth 2017}
D. Burgarth and A. Ajoy, ``Evolution-free Hamiltonian parameter estimation through Zeeman markers," {\em Phys. Rev. Lett.}, vol. 119, no. 3, 2017, Art. no. 030402.

\bibitem{burgarth 2017b}
D. K. Burgarth, ``Identifying combinatorially symmetric Hidden Markov Models," {\em Electronic Journal of Linear Algebra}, Vol. 34, pp. 393-398, 2018.

\bibitem{guta 2016}
M. Gu\c{t}\u{a} and N. Yamamoto, ``System identification for passive linear quantum systems," {\em IEEE Trans. Autom. Control}, vol. 61, no. 4, pp. 921-936, Apr. 2016.

\bibitem{levitt 2017}
M. Levitt and M. Gu\c{t}\u{a}, ``Identification of single-input-single-output quantum linear systems," {\em Phys. Rev. A}, vol. 95, no. 3, 2017, Art. no. 033825.

\bibitem{leghtas 2012}
Z. Leghtas, G. Turinici, H. Rabitz, and P. Rouchon, ``Hamiltonian identification through enhanced observability utilizing quantum control," {\em IEEE Trans. Autom. Control}, vol. 57, no. 10, pp. 2679-2683, Oct. 2012.

\bibitem{zhang 2014}
J. Zhang and M. Sarovar, ``Quantum Hamiltonian identification from measurement time traces," {\em Phys. Rev. Lett.}, vol. 113, no. 8, 2014, Art. no. 080401.

\bibitem{sone 2017}
A. Sone and P. Cappellaro, ``Hamiltonian identifiability assisted by a single-probe measurement," {\em Phys. Rev. A}, vol. 95, no. 2, 2017, Art. no. 022335.

\bibitem{sone dimension}
A. Sone and P. Cappellaro, ``Exact dimension estimation of interacting qubit systems assisted by a single quantum probe," {\em Phys. Rev. A}, vol. 96, no. 6, 2017, Art. no. 062334.

\bibitem{kato 2014}
Y. Kato and N. Yamamoto, ``Structure identification and state initialization of spin networks with limited access," {\em New J. Phys.}, vol. 16, no. 2, 2014, Art. no. 023024.

\bibitem{bellman 1970}
R. Bellman and K. J. {\AA}str\"{o}m, ``On structural identifiability," {\em Math. Biosci.}, vol. 7, no. 3-4, pp. 329-339, 1970.

\bibitem{pohjanpalo 1978}
H. Pohjanpalo, ``System identifiability based on the power series expansion of the solution," {\em Math. Biosci.}, vol. 41, no. 1-2, pp. 21-33, 1978.

\bibitem{travis 1981}
C. C. Travis and G. Haddock, ``On structural identification," {\em Math. Biosci.}, vol. 56, no. 3-4, pp. 157-173, 1981.

\bibitem{walter 1981}
E. Walter and Y. Lecourtier, ``Unidentifiable compartmental models: What to do?" {\em Math. Biosci.}, vol. 56, no. 1-2, pp. 1-25, 1981.

\bibitem{vajda 1989}
S. Vajda, K. R. Godfrey, and H. Rabitz, ``Similarity transformation approach to identifiability analysis of nonlinear compartmental models," {\em Math. Biosci.}, vol. 93, no. 2, pp. 217-248, 1989.

\bibitem{godfrey 1987}
K. R. Godfrey and J. J. DiStefano III, ``Identifiability of model parameters," {\em IFAC Proc. V.}, vol. 18, no. 5, pp. 89-114, 1985.

\bibitem{norton 1982}
J. P. Norton, ``An investigation of the sources of nonuniqueness in deterministic identifiability," {\em Math. Biosci.}, vol. 60, no. 1, pp. 89-108, 1982.






\bibitem{book 2011}
L. Wang and H. Garnier eds., {\em System Identification, Environmental Modelling, and Control System Design}. London: Springer-Verlag, 2012.














\bibitem{my 2016}
Y. Wang, D. Dong, B. Qi, J. Zhang, I. R. Petersen, and H. Yonezawa, ``A quantum Hamiltonian identification algorithm: Computational complexity and error analysis," {\em IEEE Trans. Autom. Control}, vol. 63, no. 5, pp. 1388-1403, 2018.

\bibitem{Sontag 2002}
E. D. Sontag, ``For differential equations with $r$ parameters, $2r+1$ experiments are enough for identification," {\em J. Nonlinear Sci.}, vol. 12, no. 6, pp. 553-583, 2002.

\bibitem{kalman 1963}
R. E. Kalman, ``Mathematical description of linear dynamical systems," {\em J. Soc. Ind. Appl. Math. Series A: Control}, vol. 1, no. 2, pp. 152-192, 1963.

\bibitem{distefano 1977}
J. DiStefano, ``On the relationships between structural identifiability and the controllability, observability properties," {\em IEEE Trans. Autom. Control}, vol. 22, no. 4, pp. 652-652, Aug. 1977.

\bibitem{zhou 1996}
K. Zhou, J. C. Doyle, and K. Glover, {\em Robust Optimal Control}. New Jersey: Prentice Hall, 1996.

\bibitem{my 2017}
Y. Wang, D. Dong, and I. R. Petersen, ``An approximate quantum Hamiltonian identification algorithm using a Taylor expansion of the matrix exponential function," in {\em Proc. 56th IEEE Conf. Decis. Control}, Melbourne, AUS, Dec. 12-15, 2017.

\bibitem{my ifac}
Y. Wang, D. Dong, I. R. Petersen, and J. Zhang, ``An approximate algorithm for quantum Hamiltonian identification with
complexity analysis," {\em 20th IFAC World Congress}, Toulouse, France, July 9-14, 2017.


\bibitem{gelfand 2008}
I. M. Gelfand, M. Kapranov, and A. Zelevinsky, {\em Discriminants, Resultants, and Multidimensional Determinants}. Boston: Birkh\"{a}user, 2008.



\end{thebibliography}
\end{document}